\documentclass[sigconf]{acmart}

\usepackage{graphicx}
\usepackage{float}
\usepackage{hyperref}

\newif\ifarxiv

\arxivtrue 

\newcommand{\figref}[1]{Fig.\ \ref{#1}}

\newcommand{\tabref}[1]{Tab.\ \ref{#1}}
\newcommand{\secref}[1]{Section \ref{#1}}

\copyrightyear{2023}
\acmYear{2023}
\setcopyright{rightsretained}
\acmConference[WSDM '23] {Proceedings of the Sixteenth ACM International Conference on Web Search and Data Mining}{February 27--March 3, 2023}{Singapore, Singapore.}
\acmBooktitle{Proceedings of the Sixteenth ACM International Conference on Web Search and Data Mining (WSDM '23), February 27--March 3, 2023, Singapore, Singapore}
\acmPrice{}
\acmISBN{978-1-4503-9407-9/23/02}
\acmDOI{10.1145/3539597.3570387}

\ifarxiv
\usepackage{array}
\usepackage[framemethod=tikz]{mdframed}
\definecolor{cccolor}{rgb}{.67,.7,.67}
\fi

\begin{document}
\title{Ask “Who”, Not “What”: Bitcoin Volatility Forecasting with Twitter Data}

\author{M. Eren Akbiyik}
\authornote{Authors contributed equally to the paper.}
\email{eakbiyik@ethz.ch}
\affiliation{%
  \institution{ETH Zurich}
  \streetaddress{Ramistrasse 50}
  \city{Zurich}
  \country{Switzerland}}
  
\author{Mert Erkul}
\authornotemark[1]
\email{merkul@ethz.ch}
\affiliation{%
  \institution{ETH Zurich}
  \streetaddress{Ramistrasse 50}
  \city{Zurich}
  \country{Switzerland}}

\author{Killian Kämpf}
\authornotemark[1]
\email{kkaempf@ethz.ch}
\affiliation{%
  \institution{ETH Zurich}
  \streetaddress{Ramistrasse 50}
  \city{Zurich}
  \country{Switzerland}}
  
  \author{Vaiva Vasiliauskaite}
\email{vvasiliau@ethz.ch}
\affiliation{%
  \institution{ETH Zurich}
  \streetaddress{Ramistrasse 50}
  \city{Zurich}
  \country{Switzerland}}
  
  \author{Nino Antulov-Fantulin}
\email{anino@ethz.ch}
\affiliation{%
  \institution{ETH Zurich}
  \streetaddress{Ramistrasse 50}
  \city{Zurich}
  \country{Switzerland}}
  
\begin{abstract}
Understanding the variations in trading price (volatility), and its response to exogenous information, is a well-researched topic in finance. In this study, we focus on finding stable and accurate volatility predictors for a relatively new asset class of cryptocurrencies, in particular Bitcoin, using deep learning representations of public social media data obtained from Twitter. For our experiments, we extracted semantic information and user statistics from over 30 million Bitcoin-related tweets, in conjunction with 15-minute frequency price data over a horizon of 144 days. Using this data, we built several deep learning architectures that utilized different combinations of the gathered information. For each model, we conducted ablation studies to assess the influence of different components and feature sets over the prediction accuracy. We found statistical evidences for the hypotheses that: (i) temporal convolutional networks perform significantly better than both classical autoregressive models and other deep learning-based architectures in the literature, and (ii) tweet author meta-information, even detached from the tweet itself, is a better predictor of volatility than the semantic content and tweet volume statistics. We demonstrate how different information sets gathered from social media can be utilized in different architectures and how they affect the prediction results. As an additional contribution, we make our dataset public for future research.

\end{abstract}



\maketitle

\ifarxiv
\begin{mdframed}[outerlinecolor=black,outerlinewidth=2pt,linecolor=cccolor,middlelinewidth=3pt,roundcorner=10pt]
\hspace{-4pt}\begin{tabular}{m{0.5\textwidth} m{0.4\textwidth}}
\href{http://creativecommons.org/licenses/by/4.0/}{This work is licensed under a Creative Commons Attribution 4.0 International License.} & \href{http://creativecommons.org/licenses/by/4.0/}{\includegraphics[scale=1.02]{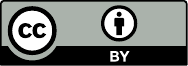}}
\end{tabular}
\end{mdframed}
\fi

\section{Introduction}
Cryptocurrency market, after its surge in 2009, gained immense popularity among not only small-scale investors, but also large hedge funds. A study in 2017 indicated that there were approximately 6 million unique users who held cryptocurrencies~\cite{crypto_bench}; in 4 years this number has risen to around 300 million~\cite{crypto_owner}. Acknowledging this exponential increase in popularity, investors began constructing portfolios using cryptocurrencies, however, the vast majority of the market share still belongs to individuals~\cite{crypto_owner}. The importance of individual investors in cryptocurrency markets incentivizes studying alternative sources of information, such as social media, to which the market participants may respond.

With its widely available API, ease of use, and its primary purpose of information sharing, Twitter is both a relevant and a convenient online resource, utilized in many past studies where the relation between financial markets and social media was investigated~\cite{tt_snp, rel_2_Bitcoin_resp, rel_3_cnn_lstm_crypto, rel_4_predict_stock_market, rel_5_influence_social_snp, rel_6_social_media_forecast, rel_7_twitter_pred_btc, rel_8_rnn, rel_9_comp_study}. Among cryptocurrencies, Bitcoin is arguably the most appropriate candidate to analyse the relationship between social media and cryptocurrency markets. Although more than 1000 cryptocurrencies are currently available, Bitcoin remains the focus of attention, likely due to its accessibility, maturity, and dominance in market capitalization.

Investigating the relationship between a financial asset and some exogenous factor requires a clear definition of the target asset's properties. Among the main aspects considered in portfolio formation~\cite{volatility_deriv}, Bitcoin's realized volatility is a property that was shown to respond to social media (in particular, Twitter) signals~\cite{rel_7_twitter_pred_btc,nino_garch}. Understanding the key drivers of interday realized volatility of cryptocurrencies is becoming more essential with the increasing popularity of high-frequency trading of Bitcoin~\cite{rise_of_machines}. Great care is required for such analysis to ensure that the learnt model is causal and robust under exogenous perturbations for the out-of-sample predictions. 

Building upon these insights, our goal in this study is to determine which components of the publicly available Twitter data have explanatory capacity for predicting daily realized volatility of Bitcoin prices. We note that we do not aim to find the state-of-the-art model that can predict the price volatility of Bitcoin over some test dataset. Instead, we intend to systematically compare the classical financial models with deep learning-based approaches for volatility forecasting, and to understand the key exogenous (social media) features or representations that may influence realized volatility, when combined with endogenous market features. To achieve this goal, we utilize a modular and testable architecture to which social media and other features can be progressively added. 

For the prediction task, we considered the intraday logarithmic returns, along with publicly available Twitter data, such as tweet counts over 15-minute time frames. Furthermore, we extracted semantic information and user interaction statistics from over 30 million Bitcoin-related tweets. The text content of tweets has to be “interpreted”, i.e., assigned emotional sentiment, or some other machine-comprehensive meaning. Here we used Valence Aware Dictionary for Sentiment Reasoning (VADER)~\cite{vader}, a well-known natural language processing tool for extracting emotional sentiment of textual data. Noting that certain players in both cryptocurrency market and in social media are more influential than others, we also incorporated statistics of user accounts, such as follower count, verified status, and activity to our feature space. This allowed us to separately examine the importance of individual tweets and the potential influence of notable agents. 
To perform a comparative analysis of different feature sets, we designed a deep learning architecture that can flexibly make use of different feature spaces. We conducted ablation studies to statistically compare the information content of different feature sets, while controlling for the added complexity of model architectures and parameter spaces. We first showed that the deep learning models are promising tools for Bitcoin volatility forecasting, outperforming classical financial forecasting methods. We then upgraded the most promising deep learning model to allow for Twitter-related features. Our results suggested that certain (but not all) features of social media are accurate and useful predictors of inter-day Bitcoin volatility. 

We designed the study such that the model architectures are preserved under changes in feature sets and data formats. In this way, a statistically significant and efficient comparison is possible. To encourage the extension of this study via more sophisticated analyses that are outside the scope of the current work, e.g., of possible interaction effects of feature sets, or employing more complex neural models such as Graph Neural Networks, we are making code and dataset public in \href{https://github.com/meakbiyik/ask-who-not-what}{github.com/meakbiyik/ask-who-not-what}.

This paper is organized as follows. In \secref{sec:related_work}, we discuss the previous literature. In \secref{sec:data} we present the data collection process, followed by a definition of a volatility prediction task in \secref{sec:prediction}. In \secref{sec:tcn} and \secref{sec:twitter_tcn}, we describe classical econometric models and study deep learning architectures that incorporate various amounts of price and social media information. We conclude and discuss the implications of our study and results in \secref{sec:discuss}.

\section{Related Work}\label{sec:related_work}

Historically, asset prices dynamics~\cite{stanley2000introduction} was modelled with a Wiener process or Brownian motion. Although Brownian motion can explain certain empirical facts of price change dynamics~\cite{cont2001empirical}, others, such as volatility clustering, cannot. 
Autoregressive conditional heteroscedasticity (ARCH) model~\cite{ARCH} was introduced to describe the time-varying volatility of logarithmic price returns, relying only on the information of previous price movements. Its generalized variant GARCH~\cite{GARCH} additionally introduces previous conditional variances when calculating the present conditional variance. GARCH is thus able to explain the heavy-tail distribution of price returns and volatility clustering.
The effect of news on price volatility was modelled with the class of (conditional) jump models~\cite{jorion1988jump,chan2002conditional}, incorporate Poisson-like components.

We also note that models based on the autoregressive realized volatility~\cite{ARCH_HAR} appear to be more precise in forecasting volatility than conditional variance models. 

State-of-the-art time series analysis tools, machine learning models and deep learning models have been extensively used in financial contexts in recent years. 
Recurrent Neural Networks (RNNs) and, specifically, Long Short-Term Memory networks (LSTMs)~\cite{liu2019novel,nguyen2019long} were shown to be reliable and informative for a diverse set of financial tasks~\cite{rel_9_comp_study}. Lately, convolutional neural networks (CNNs) for causal temporal analysis, also known as Temporal Convolutional Networks (TCNs), have also gained traction~\cite{mcnally2018predicting,saad2019toward}. Models that combine these two paradigms were also studied~\cite{guo2021mrclstm,rel_3_cnn_lstm_crypto}.
The empirical studies that analyse drawbacks and advances of volatility forecasting with neutral networks are still under active research~\cite{bucci2020realized, vejendla2013evaluation}. 

Stylized facts are not fully known in cryptocurrency markets, and it is not entirely clear what the economic value of cryptocurrencies is~\cite{Hayes2015,Bolt2016,nadarajah2017inefficiency}, which econometric models are relevant, and how price discovery~\cite{cheah2015speculative,kristoufek2015main,dimpfl2021nothing} or bubbles~\cite{BouchaudBTC} occur. Therefore, cryptocurrency markets have been studied from various alternative perspectives, ranging from volatility and volume forecasting using standard econometric models~\cite{Katsiampa2017,GuoBifetAntulov18,antulov2021temporal}, to employing tools from systems dynamics~\cite{Ron2013BTC,ElBahrawy2017,AntTol18}.
Various studies analysed the effect of external data from social media, news, search queries, sentiment, comments, replies on forums, and blockchain~\cite{Garcia2015, DeyBlock2020, nino_garch,beck2019sensing}. Typically, sentiment analysis is performed and combined with econometric models such as GARCH~\cite{rel_5_influence_social_snp} or deep learning models such as CNN~\cite{rel_2_Bitcoin_resp}. 
The impact of social media and other exogenous factors on price in both cryptocurrency and conventional markets suggests that prices respond to more than price history. The inclusion of data from social media, news feed or blockchain to improve forecasts has been studied widely~\cite{rel_9_comp_study,rel_5_influence_social_snp,rel_6_social_media_forecast,rel_7_twitter_pred_btc}. While it seems the impact of positive and negative sentiment is different~\cite{epstein}, literature suggestions are inconclusive and potentially case-dependent. Finally, it is worth mentioning the class of Hawkes models~\cite{cao2017deephawkes,mei2017neural} have been used for the quantification of ``self-exciting'' processes of retweet dynamics on social media.

\section{Data Pipeline}\label{sec:data}

\subsection{Data Collection}
The data collection pipeline consists of two different components, namely of the tweet data and of the Bitcoin price data.

\subsubsection{Tweet data} 
Using the publicly available Twitter API, the tweets were acquired via a real-time streamer (\emph{stream-watcher}). All \emph{relevant} tweets between 10.10.2020 and 3.3.2021 were stored as \emph{JSON line objects} on an AWS MongoDB server. A tweet was considered relevant, if its textual body contained one of the following strings: ``BTC'', ``\$BTC'', or ``Bitcoin''. Such a tweet was automatically appended to the database by the streamer.

\subsubsection{Price Data}
We also considered 15-minute closing prices in Bitfinex exchange for the prediction task, since this exchange provides accurate high-frequency data free of charge as opposed to its alternatives. The data was acquired using 
\href{https://github.com/ccxt/ccxt}{Cryptocurrency eXchange Trading} Library.

Overall, the dataset used in this study consisted of approximately 14000 Bitcoin price snapshots and 30 million Bitcoin-related tweets.
\subsection{Data Preprocessing} \label{sec:preprocess}
The collected data was preprocessed in order to reduce the memory footprint and homogenize the dataset schema.

\paragraph{Refactoring}
A JSON line object for a single tweet consists of nested objects that depend on a tweet's type: a \emph{general} tweet, a \emph{quote} tweet, a \emph{retweet}, or a \emph{reply} to a general tweet. Replies, quotes, and retweets include an additional data entry that contains the original tweet's content. Therefore, these types of tweets may be larger in size than general tweets. To make all types of tweets equivalent in our database and to reduce overall redundancy, the fields, such as media and retweeted tweet's body, were converted to flags or counts, so that the refactored structure for all types of tweets is identical.

\paragraph{Pruning}
After the refactorization, the JSON line object consists of approximately 120 fields, including information about its location and source (i.e., URLs for media-including tweets). For the study, some of these fields were redundant, therefore they were excluded. The selected fields could be partitioned into \emph{user-related} and \emph{tweet-related} fields. The schema of the pruned and refactored tabular format is shown in \tabref{tab:data_schema}.

\paragraph{User-related fields}
Assuming that some data related to the author of the tweet could be informative, we considered parts of it. In particular, hypothesizing that popular accounts may have large influence, we used data fields that relate to the popularity of a user. 

\paragraph{Tweet-related fields}
The previously mentioned tweet streamer collects the sample tweet upon its creation. At the time, retweet count, favourites count, and quote count are zero, thus, they would not aid the predictions. Acknowledging this, we selected tweet-related fields that describe the tweet's content rather than its popularity. These fields include the number of media objects in the tweet, e.g.\ whether it is a quote tweet or a retweet, whether it contains sensitive language, and, finally, the main body of the tweet.

After refactoring and pruning, the entire dataset was sorted and indexed by JSON line objects' creation timestamps, after which the dataset was stored in a Parquet format. This pipeline ensured the tweets are stored in a tabular manner (rather than a tree structure), making columnar processing, compression/decompression efficient.

\begin{table}[!ht]
  \caption{Pruned and refactored JSON object schema of the preprocessed Twitter data.}
  \label{tab:data_schema}
\resizebox{0.45\textwidth}{!}{%
    \begin{tabular}{@{}llll@{}}
    \toprule
    \multicolumn{2}{c}{\textbf{User-related}} & \multicolumn{2}{c}{\textbf{Tweet-related}} \\ 
    \cmidrule(lr){1-2} \cmidrule(lr){3-4}
    \multicolumn{1}{c}{Name} & \multicolumn{1}{c}{Data type} & \multicolumn{1}{c}{Name} & \multicolumn{1}{c}{Data type} \\ 
    \cmidrule(lr){1-1} \cmidrule(lr){2-2} \cmidrule(lr){3-3} \cmidrule(lr){4-4}
    \textsc{favourites\_count} & integer & \textsc{created\_at} & timestamp \\
    \textsc{followers\_count} & integer & \textsc{gif\_count} & integer \\
    \textsc{friends\_count} & integer & \textsc{photo\_count} & integer \\
    \textsc{listed\_count} & integer & \textsc{video\_count} & integer \\
    \textsc{verified} & boolean & \textsc{is\_quote\_status} & boolean \\
    \textsc{default\_profile} & boolean & \textsc{possibly\_sensitive} & boolean \\
    \textsc{default\_profile\_image} & boolean & \textsc{tweet\_text} & string \\ 
    \bottomrule
    \end{tabular}
    }
\vspace{-10pt}
\end{table}

\subsection{Semantic Embedding of Tweet Text}

Each tweet's text (\textsc{tweet\_text}) was fed through a well-known sentiment tool, Valence Aware Dictionary for Sentiment Reasoning (VADER)~\cite{vader}, implemented in \href{https://www.nltk.org/api/nltk.sentiment.vader.html}{NLTK}. VADER was used to obtain an impression of tweet author's stance towards the cryptocurrency at the time of tweeting (\textsc{created\_at}). One of the several advantages that VADER provides with is multilingualism, therefore the whole corpora of tweets need not be translated into one language. The VADER output, a four-dimensional vector, contains sentiment scores for positive, neutral, and negative emotions, and the compound score. The compound score incorporates the three raw sentiment scores via a heuristic function, whose output is normalized to range between $[-1, 1]$. We used the compound score instead of the raw main body text to represent the textual content of the tweet in the forecasting model.

\section{Prediction Task}\label{sec:prediction}

Incorporating semantic information, tweet and user metadata into a single realized volatility model is not straightforward. Although numeric data can be directly employed via, e.g., linear, autoregressive models~\cite{time_series} given the stationarity of the inputs, the task is more challenging when multiple data types are considered, such as text and other media.

\paragraph{Target} We aim to determine whether Twitter data, alongside with log-returns, can be used to predict the Bitcoin price volatility. Log-returns are defined as the change in logarithm of price as observed at two consecutive time steps, namely,
\begin{equation}
    r_t = \log{P_t} - \log{P_{t-1}} .
\end{equation}
\emph{Realized volatility} is an empirical measure of return variability, defined as the square-root of summed squared log-returns in a time period $\Delta t$~\cite{realized_volatility,ARCH_HAR}:
\begin{align}
    RV_{\Delta t} &= \sqrt{\sum_{t \in \Delta t} r_t^2}. \label{eq:realized_volatility}
\end{align}

The \emph{target} for all models is the next day's realized volatility, obtained from 96 segments of 15-minute log-return predictions, and aggregated using standard formula \eqref{eq:realized_volatility}.

\paragraph{Evaluation Metrics} We considered a set of standard metrics to compare the real $RV$ with the predicted volatility, and to provide a single score that displays the quality of the model under certain assumptions. We used the following metrics: \emph{mean absolute percentage error} (MAPE), \emph{mean absolute error} (MAE), \emph{root-mean-square error} (RMSE), and \emph{mean squared logarithmic error} (MSLE). Their comparison provides an insight on the behaviour of the predicted values under varying conditions of the true signal. 

\section{Volatility Forecasting}\label{sec:tcn}

As a first step, we studied which deep learning architectures may surpass the classical financial models. We use a general setup of forecasting models, where historic data is used to predict the next value of a target variable. Typically, volatility is predicted based on feature vectors that relate to past price returns, e.g., as is done in GARCH~\cite{nino_garch}, or to past volatility estimates, as in (H)AR-RV~\cite{ARCH_HAR}. In this part of the study, we considered GARCH and AR-RV models as \emph{classical econometric baselines}, and compared their performance to \emph{deep learning models} that utilize the same financial information.

\subsection{Classical Econometric Baselines}

As described briefly in \secref{sec:related_work}, GARCH models are generalized versions of ARCH models, incorporating previous conditional variances ($\sigma^{2}_{t-n}$) when calculating the present conditional variance, as well as the previous squared logarithmic returns ($r^{2}_{t-m}$). Specifically, the mathematical formula for GARCH$(m, n)$ is defined as:
\begin{align}
\sigma^{2}_{t} = \alpha_{0} + \alpha_{1}r^{2}_{t-1} + ... + \alpha_{m}r^{2}_{t-m} + \beta_{1}\sigma^{2}_{t-1} + ... + \beta_{n}\sigma^{2}_{t-n}.
\end{align}
On the other hand, (H)AR-RV models the realized volatility using the realized volatility as input. Heterogeneity (H) is achieved by different horizons for realized volatility, namely daily ($d$), weekly ($w$) and monthly ($m$). The HAR-RV model is defined as follows:
\begin{align}
RV^{d}_{t+1} = \alpha_{0} + \alpha_{d}RV^{d}_{t} + \alpha_{w}RV^{w}_{t} + \alpha_{m}RV^{m}_{t}.
\end{align}

In consideration of the available data format and the determined time horizon of analysis, we only inputted the daily RV values with 5-day lags, making our classical econometric baseline a standard AR-RV without heterogeneous horizons.

\subsection{Time-Series Forecasting with Deep Learning} RNN-based models are popular for financial applications. They have also been successfully explored as models that utilize Twitter data to analyse the cryptocurrency market~\cite{rel_3_cnn_lstm_crypto, rel_9_comp_study}. However, convolutional networks have consistently shown similar or better results in a variety of tasks with fewer number of parameters~\cite{RNNvsTCN}. RNNs also seem to have a shorter effective memory than their convolutional counterparts~\cite{RNNvsTCN}. Despite theoretically infinite receptive fields, RNNs suffer from inherent exploding/vanishing gradient problems, and are susceptible to neuron saturation because of the sigmoid activations. RNN-based models have different variations that utilize gating mechanisms to extract more sophisticated and long-term information from the data, but the most appropriate variation is task-dependent~\cite{chung2014gru}.

In order to select the most appropriate deep learning architecture that can be further expanded to incorporate social media data in the incoming tasks, we first compared LSTM, GRU and TCN using solely the price data as the input. As the autoregressive models have already proven to be remarkably successful in volatility forecasting~\cite{ARCH_HAR}, we expect deep architectures to at least perform as well as these baselines.

\begin{figure}[!bth]
    \centering
    \includegraphics[width=0.45\textwidth]{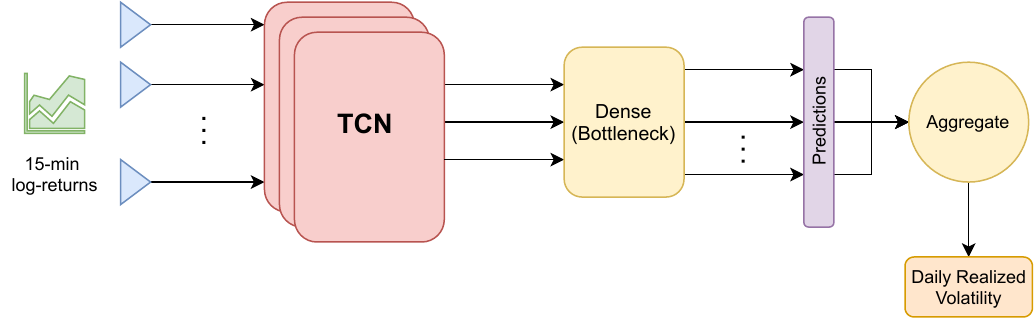}
    \caption{Baseline TCN model. Causal convolution over time with bottleneck interpolator. For RNN-based models, the TCN block is replaced by LSTM or GRU block.}
    \label{fig:tcn_baseline}
\vspace{-10pt}
\end{figure}

RNN-based models and TCN-based models that were considered in this study utilize similar design, illustrated in \figref{fig:tcn_baseline}. The receptive field of all models is fixed, and the inputs are directly fed to the respective time-series models without any preceding layers. Both input and output features are in log-return space, and the metric scores are calculated after aggregating the 15 minute-level output predictions for a day to estimate the daily volatility using \eqref{eq:realized_volatility}. The losses are directly estimated from observed log-returns (rather than realized volatilities) to provide more fine-grained gradient estimates to the network.

\paragraph{TCN mechanism} TCNs were introduced to capture temporal information using 1-dimensional causal convolutions that avoid information leakage from the future time steps~\cite{tcn_paper, stcn}. However, without the reduction of dimensionality in subsequent layers, convolution operation becomes very costly for long input signals. To bypass this potential limitation, dilated convolutions for long-history dependent sequence tasks were proposed~\cite{wavenet, RNNvsTCN}. For a 1-dimensional signal $\textbf{x}$ at time $t$ of a sequence $s$ to be mapped to values $\textbf{y}$, and a filter $f$, one can define the dilated convolution operation $F$ as follows:
\begin{equation}
\label{eq:dil_conv}
    F(s) = (\textbf{x} *_d f)(s) = \sum_{i=0}^{k-1}f(i)\textbf{x}_{s-d\times i}
\end{equation}
with $d$ the dilation factor, $k$ the kernel (filter) size, and $s-d\times i$ accounting for the signals acquired earlier in the sequence~\cite{RNNvsTCN}. It is important to note that for $d=1$, \eqref{eq:dil_conv} is equal to a standard 1-dimensional convolution.

\paragraph{Bottleneck layer} 
Noting that encoder-decoder architectures were more prone to overfitting on high-frequency artefacts in our data, we relied on a simpler decoding scheme with a linear layer. 
This design choice ensured the model focuses more on the low-frequency components of the output signal while providing the desired output variability. Another potential explanation for the effectiveness of this approach is that the linear interpolation estimates intraday stationarity at no cost, therefore allowing the TCN to train on variance-inducing components of the input data.

\paragraph{Loss function} Presence of quasi-linear changes in price suggests that the target vector mainly consists of small and analogous values of log-returns.  Mean squared error loss, which is quite often used in the regression tasks in training neural networks, is observed to trap the network at a local minimum of constant output as small log-returns dominate the weight gradients and prevent the network from learning about the meaningful changes. To resolve this issue, we resorted to \emph{squared epsilon-insensitive} loss function that ignores the prediction error when the value is closer than epsilon $\epsilon$ to the ground truth:
\begin{align}
\mathcal{L}_\epsilon(r_t, \hat{r}_t)
= \max\{0, (r_t-\hat{r}_t)^2 - \epsilon\}.
\end{align}
Here $r_t$ stands for the log-return at the time point $t$, $\hat{r}_t$ is the prediction of the model for the same value, and $\epsilon$ is the upper end of the interval $[0, \epsilon]$ in which the loss is ignored.  

\begin{table*}[!thb]
  \caption{Performances of the autoregressive baselines and deep learning models across 20 runs.}
  \label{tab:baseline_res}
  \centering
  \resizebox{0.5\textwidth}{!}{
  \begin{tabular}{lllll}
    \toprule
     &\textbf{MAPE} &\textbf{MAE} &\textbf{RMSE} &\textbf{MSLE}\\
    \midrule
    \text{AR-RV} & 0.32 & 1.91 & 2.78 & 0.12 \\
    \text{GARCH} & 0.35 & 2.21 & 2.93 & 0.14 \\
    \hline
    \text{LSTM} & 0.30   ± 0.08 & 1.70   ± 0.26  ** & \textbf{2.37   ± 0.19 ***} & \textbf{0.09  ± 0.01 ***}\\
    \text{GRU} & 0.57   ± 0.40 & 2.92  ± 2.3 & 4.23   ± 5.61 & 0.20   ± 0.20 \\
    \textbf{\text{TCN}} & \textbf{0.23   ± 0.01 ***} & \textbf{1.63   ± 0.06 ***} & 2.58   ± 0.07 *** & 0.11   ± 0.01 *** \\
    \bottomrule
  \end{tabular}
  }
  \medskip\\
\textbf{For the deep learning models:} One-sided t-test significances with respect to AR-RV \\
\textbf{Significance codes:} 0  - ‘***’ - 0.001 - ‘**’ - 0.01 - ‘*’ - 0.05 - ‘.’ - 0.1 
\vspace{-10pt}
 \end{table*}
\subsection{Experiment Setup}\label{sec:exp_setup}

The experimental setup for the autoregressive models and the deep learning architectures have a common training horizon of 96 days and a testing horizon of 48 days. AR-RV model uses daily RV values as inputs, whereas GARCH model uses log-returns. The input data is scaled to reside between $[-0.25, 0.25]$ by subtracting the minimum from the data and dividing the result with the data range, extracted from the training set. In all subsequent experiments, daily realized volatility is predicted using the available information from the previous day, and training is conducted in 1-day strides.

\paragraph{Hyperparameter optimization} Deep learning models often suffer from bad hyper-parametrization and the determination of such parameters can be a critical source of bias. In order to prevent any imbalance of attention from affecting the results of the experiment, we conducted thorough hyperparameter searches for each model with predetermined resources. Given the $72-24$ day split of the training data into train and validation sets, all three models were tuned to minimize MAPE metric with a multivariate Bayesian optimizer for 250 different configurations using Python library \textit{optuna}~\cite{optuna}. The optimized parameters include the individual components of each model, such as recurrent unit count and kernel size, as well as common parameters, such as the learning rate and the loss epsilon $\epsilon$. The epoch count is fixed at 30 for all configurations. 
All experiments were conducted on Nvidia Titan X GPUs with Pascal architecture for the deep learning models. Autoregressive baseline models were evaluated using CPU. Each non-deterministic model was trained 20 times, and the results were compared using appropriate one-sided statistical tests. For all models, we used AdamW optimizer~\cite{adamw} with optimal learning rate and weight decay as found by the hyperparameter optimization.

\begin{figure}[!tbh]

\begin{minipage}[b]{0.48\linewidth}
  \centering
  \centerline{\includegraphics[width=\linewidth]{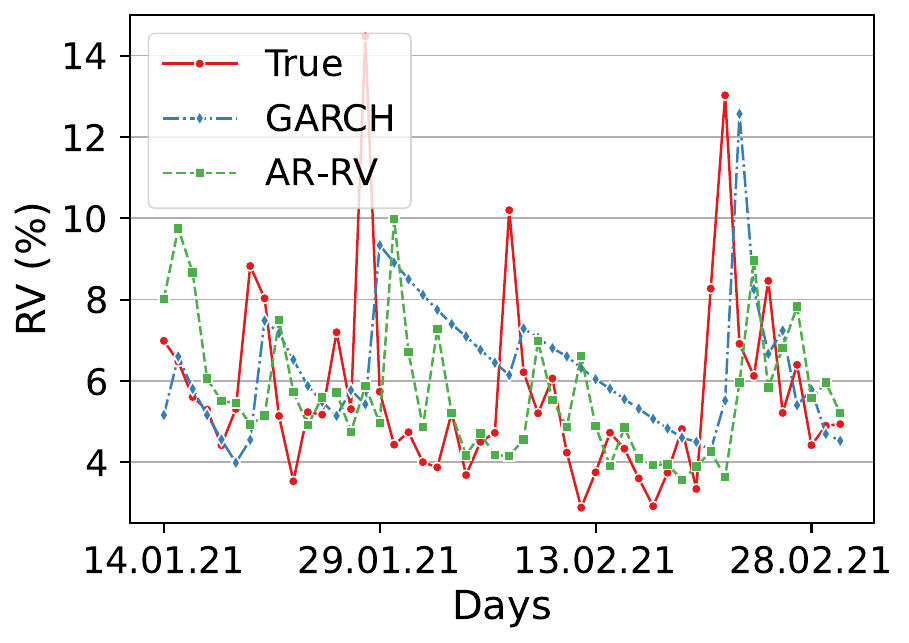}}
  \centerline{(a) AR-RV and GARCH}
\end{minipage}
\hfill
\begin{minipage}[b]{0.48\linewidth}
  \centering
  \centerline{\includegraphics[width=\linewidth]{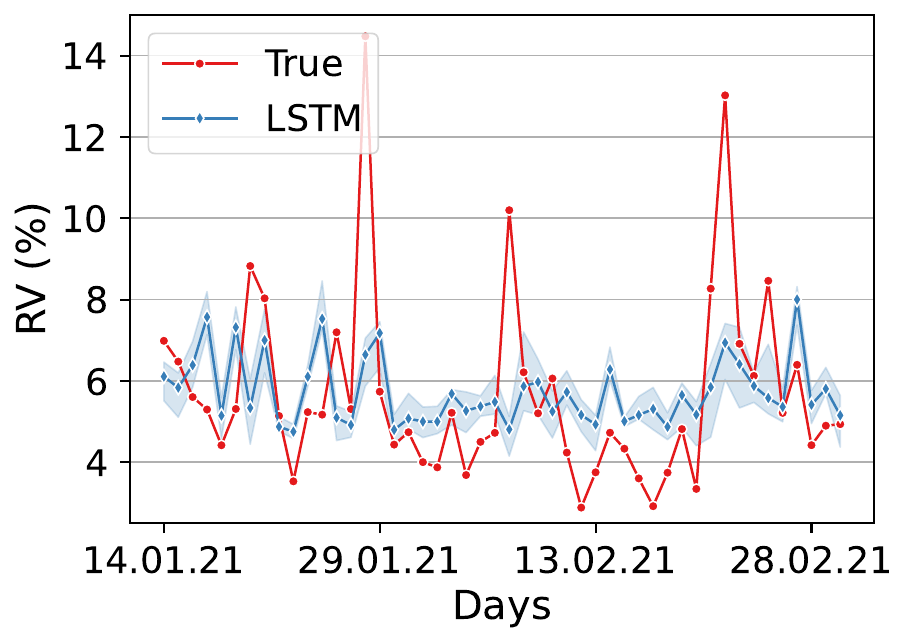}}
  \centerline{(b) LSTM}
\end{minipage}

\bigskip

\begin{minipage}[b]{0.48\linewidth}
  \centering
  \centerline{\includegraphics[width=\linewidth]{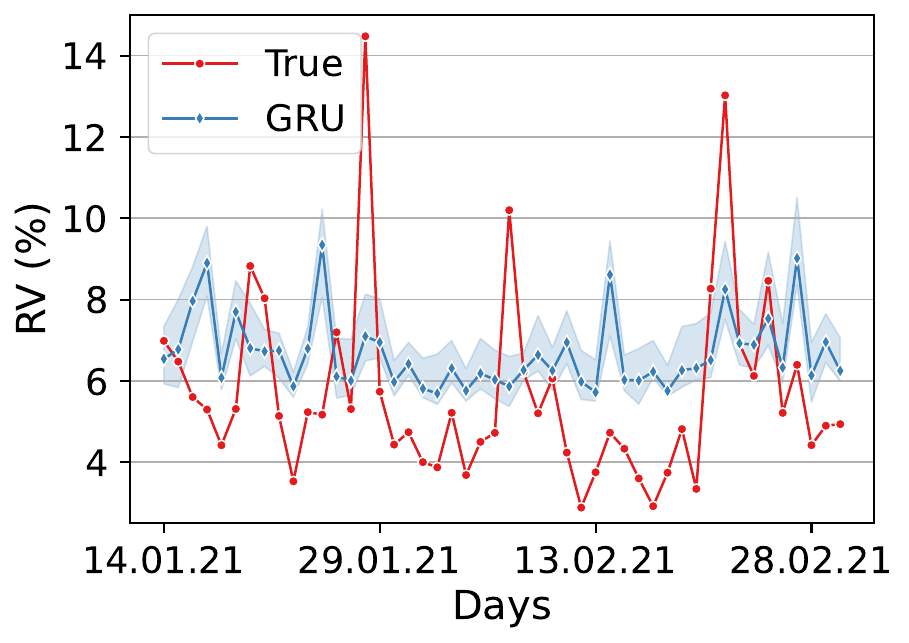}}
  \centerline{(c) GRU}
\end{minipage}
\hfill
\begin{minipage}[b]{0.48\linewidth}
  \centering
  \centerline{\includegraphics[width=\linewidth]{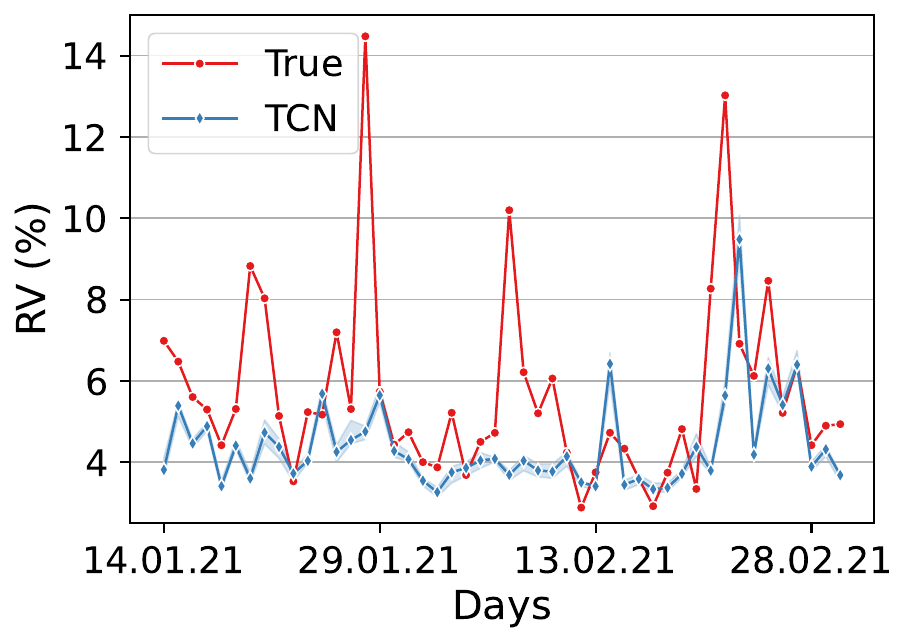}}
  \centerline{(d) TCN}
\end{minipage}
\caption{Comparison of the ground truth test time series and predictions obtained using a) AR-RV and GARCH, b) LSTM, c) GRU, and d) TCN. For subfigures b-d, shaded blue areas indicate 95\% confidence intervals, estimated via bootstrapping (using 1000 bootstrap samples).}
\label{fig:baselines}
\vspace{-10pt}
\end{figure}

\subsection{Results}

We compared the performances of the econometric baselines, with TCN, GRU, and LSTM over the 48 days testing horizon. Here we primarily focus on MAPE metric, as it ensures that we do not inflate errors due to high-volatility data points (such high-volatility incidents in Bitcoin data are common and are often caused by exogenous and highly unpredictable events).

The autoregressive baseline results in \tabref{tab:baseline_res} indicate the superiority of AR-RV over GARCH based on all analysed metrics, therefore we selected AR-RV as the baseline to compare the deep learning models to. We conducted one-sided t-tests with respect to the AR-RV results to test if the mean of an error for a given model is statistically different from the error achieved by the AR-RV.

\tabref{tab:baseline_res} shows that TCN model significantly outperforms the baseline AR-RV in all metrics, whereas GRU performs worse than both of the autoregressive models. LSTM is not significantly better in terms of MAPE and shows high levels of standard deviation in different runs. Among the deep learning models, TCN is also superior in more outlier-robust metrics as MAPE and MAE, yet LSTM appears to outperform TCN in MSLE and RMSE. However, it should also be noted that MSLE is not a symmetric metric, as under-predictions are punished more than over-predictions, and might mislead in comparing the accuracy of the two models.

In \figref{fig:baselines}, we show predictions obtained by each model for a 48-day test set. Based on the confidence intervals achieved by the neural network models, we observe that TCN model is much more stable than RNN-based models. TCN consistently provides better estimates on low-volatility days. The figure also suggests that RMSE performance of LSTM can be due to overall higher mean of its predictions that improves the resulting performance metrics at the extreme data points, such as the three prominent peaks at 28.01, 07.02 and 22.02.

\begin{figure}[!bht]

\includegraphics[width=0.75\linewidth]{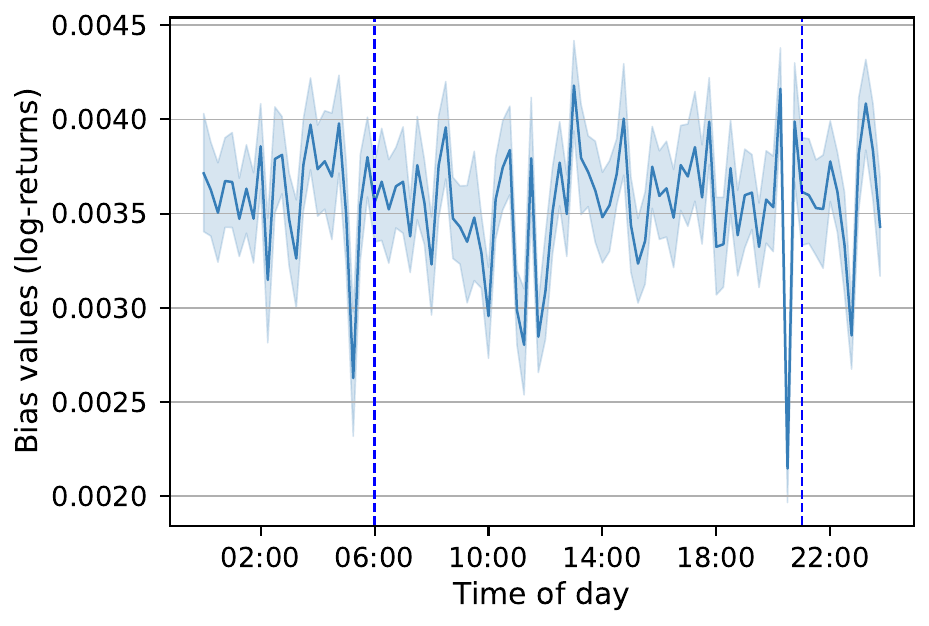}
\caption{TCN bottleneck layer bias distribution (actual scale). The values are in log-return space. Time of day is reported in UTC.}
\label{fig:bias_dist}
\vspace{-10pt}
\end{figure}

\paragraph{Bias distribution of the bottleneck layer} The final layer of the TCN, as shown in \figref{fig:tcn_baseline} explicitly models the output distribution in a way that can be visualized as the expected log-return value for each 15-minute period in one day. The distribution of the bias values, shown in \figref{fig:bias_dist}, appears stationary, with two significant downward spikes at 5:00 and 20:30. In an attempt to explain these phenomena, we have annotated the market closing times of the two countries, USA and Japan, that houses three biggest stock markets in the world, namely NYSE, NASDAQ and Japan Exchange Group. However, we do not believe that the explanation of these artefacts falls within the scope of our study, and invite the researchers to analyse or discard these.

Following the results of the experiments, \textbf{we conclude that TCN is a promising candidate model for analysis and prediction of Bitcoin volatility}. Therefore, in the following section where we also consider Twitter data, we build our models upon the TCN base model.

\section{Incorporating Twitter Data with TCN} \label{sec:twitter_tcn}

\begin{figure}[!bth]
    \centering
    \includegraphics[width=0.45\textwidth]{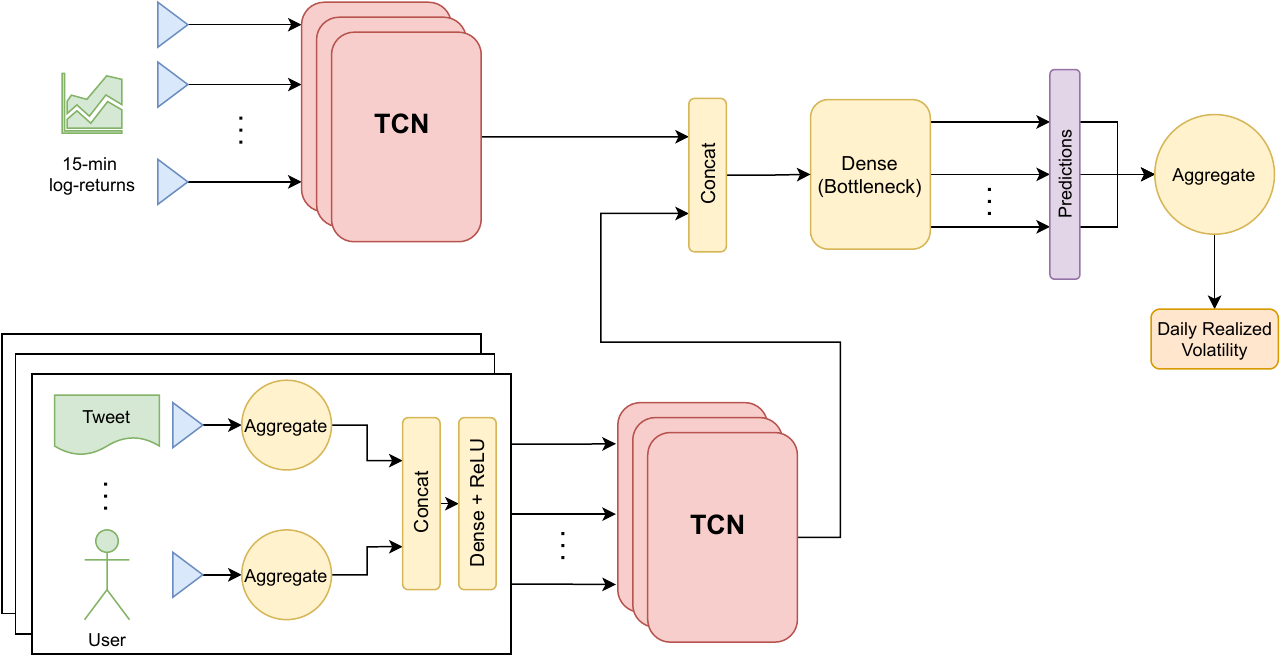}
    \caption{D-TCN model. A modular architecture with two temporal convolutions and flexible concatenation layers.}
    \label{fig:dtcn}
\vspace{-5pt}
\end{figure}

A statistical assessment of the effectiveness of Twitter data requires control of all other factors that might induce additional variance. For this purpose, we are proposing a modular architecture with two temporal convolutions and flexible concatenation layers, as depicted in \figref{fig:dtcn}. This model can be used to perform log-return predictions for different sets of data with minimal change to parameter count at and after the input layer. We call this model \emph{Double-TCN} (D-TCN), for the usage of two fully separate TCN pipelines in the architecture.

The upper TCN pipeline is identical to the model illustrated in \figref{fig:tcn_baseline}, and, as before, its purpose is to utilize the previous day's log-return values. The lower TCN pipeline, on the other hand, analyses any additional numeric data obtained from tweets. The preprocessed Twitter data (see \secref{sec:preprocess}) is aggregated to 15-minute intervals that correspond to the 15-minute intervals of log-returns. The data is then fed into a Dense+ReLU stack before the TCN. Regardless of the number of input features, the input dimension of the TCN layer is fixed across all experiments.

\paragraph{Single versus double TCN} We propose this D-TCN design both for performance and comparability: removal and addition of features is easy, and done in a way that preserves the dimensions in the rest of the network. On the contrary, in a single TCN structure the log-returns and Twitter data are concatenated and fed to the same temporal convolutional network. This may lead to a sharp performance drop, likely because of inherent noise of the tweet-based data. Therefore, a prior dimensionality reduction is required to gather meaningful evidence of volatility, instead of a further parametric expansion of the network. Furthermore, comparison of different features become an intractable task, as the network itself changes drastically to the level that the information passed down from the log-returns becomes inseparable from the tweet-based data.

\subsection{Feature Sets}
We split the pruned and refactored tweet data features indicated in \tabref{tab:data_schema} into \textbf{\textit{Tweet}}, \textbf{\textit{User}} and \textbf{\textit{VADER}} feature sets (the VADER compound score of the \textsc{tweet\_text} from \tabref{tab:data_schema}). Additionally, we aggregated the tweet counts within each 15-minute interval as \textbf{\textit{Count}} feature. 
 The split for \textbf{\textit{Tweet}} features and \textbf{\textit{User}} features is listed in \tabref{tab:feature_sets}. All but the \textbf{\textit{Count}} feature are the averages of the tweets in their respective intervals.

\begin{table}[!tbh]
    \caption{Feature sets of Twitter data}
    \label{tab:feature_sets}
    \centering
    \resizebox{0.45\textwidth}{!}{
    \begin{tabular}{llll}
    \toprule
      \textbf{\textit{Tweet}} & \textbf{\textit{User}}& \textbf{\textit{VADER}}&\textbf{\textit{Count}} \\      
        \hline
        \textsc{gif\_count } & \textsc{favourites\_count} & VADER & Total\\
        \textsc{photo\_count} & \textsc{followers\_count} & compound&number  \\
       \textsc{video\_count} & \textsc{friends\_count } & score &of tweets\\
       \textsc{is\_quote\_status} & \textsc{listed\_count } &\\
      \textsc{possibly\_sensitive }& \textsc{verified} &\\
         & \textsc{default\_profile} &\\
        & \textsc{default\_profile\_image}&\\
        \multicolumn{3}{c}{\upbracefill} & \\
         \multicolumn{3}{c}{Averaged over set of tweets in the interval} & \\
    \bottomrule
    \end{tabular}
    }
\vspace{-10pt}
\end{table}

\paragraph{Tweet features}
The tweet features in our feature set use media information contained in the tweet by enumerating the number of objects contained in the tweet body, such as GIFs, photos, and videos. \textsc{is\_quote\_status} is the saved tweet being a reply, retweet, or quote retweet of a different tweet, and \textsc{possibly\_sensitive} indicates if the tweet contains sensitive language or not.

\paragraph{User features}
We used favourites count (number of tweets the user has marked as favourite), followers count (number of followers the user has), friends count (number of accounts the user follows), listed count (number of lists the user is a member of), verified (whether the user is verified or not) and finally default profile flags indicating whether the user defined his own profile image and his own background image, or is using the default placeholders set by Twitter.

\begin{table*}[!thb]
  \caption{Performances of D-TCNs with different feature sets across 40 runs.}
  \label{tab:results}
  \centering
  \resizebox{0.65\textwidth}{!}{
  \begin{tabular}{lllll}
    \toprule
     &\textbf{MAPE} &\textbf{MAE} &\textbf{RMSE} &\textbf{MSLE}\\
    \midrule
    \text{TCN} & 0.23   ± 0.01 & 1.63   ± 0.06 & 2.58   ± 0.07 & 0.11   ± 0.01 \\
    \hline
    \textbf{\text{$\text{D-TCN}_\text{User}$}} & \textbf{0.20   ± 0.01 ***} & \textbf{1.44   ± 0.03 ***} & \textbf{2.36   ± 0.05 ***} & \textbf{0.08   ± 0.01 ***}\\
    \textbf{\text{$\text{D-TCN}_\text{Tweet}$}} & \textbf{0.21   ± 0.01 ***} & \textbf{1.59   ± 0.08 ***} & \textbf{2.45   ± 0.09 **} & \textbf{0.09   ± 0.01 **} \\
    \text{$\text{D-TCN}_\text{VADER}$} & 0.26   ± 0.02 & 1.72   ± 0.08 & 2.70  ± 0.09 & 0.13   ± 0.01  \\
    \text{$\text{D-TCN}_\text{Count}$} & 0.32   ± 0.05 & 1.97   ± 0.19  & 3.27   ± 0.47 & 0.16   ± 0.02 \\ 
    \hline
    \textbf{\text{$\text{D-TCN}_\text{VADER, Tweet, User}$}} & \textbf{0.21 ± 0.01 ***} & \textbf{1.47 ± 0.05 ***} & \textbf{2.38 ± 0.06 ***} & \textbf{0.09 ± 0.01 ***} \\
    \text{$\text{D-TCN}_\text{VADER, Tweet}$} & 0.22   ± 0.01  & 1.56   ± 0.09  & 2.51   ± 0.09  & 0.10   ± 0.01  \\
    \text{$\text{D-TCN}_\text{VADER, User}$} & 0.22 ± 0.01 ** & 1.51 ± 0.05 *** & 2.43 ± 0.09 *** & 0.10 ± 0.01 ** \\
    \text{$\text{D-TCN}_\text{Tweet, User}$} & 0.21   ± 0.01 *** & 1.49   ± 0.06 *** & 2.41   ± 0.08 *** & 0.09  ± 0.01 *** \\
    \textbf{\text{$\text{D-TCN}_\text{Count, Tweet, User}$}} & \textbf{0.22   ± 0.01 ***} &\textbf{ 1.46   ± 0.05 ***} & \textbf{2.36   ± 0.10 ***} & \textbf{0.08   ± 0.01 ***} \\
    \text{$\text{D-TCN}_\text{Count, Tweet}$} & 0.23   ± 0.01 & 1.59   ± 0.08  & 2.56   ± 0.11  & 0.11   ± 0.01  \\
    \text{$\text{D-TCN}_\text{Count, User}$} & 0.22   ± 0.01  & 1.53   ± 0.08 * & 2.46   ± 0.11 * & 0.10   ± 0.01 \\
    \text{$\text{D-TCN}_\text{Count, VADER, Tweet, User}$} & 0.22   ± 0.01 * & 1.48   ± 0.07 *** & 2.43   ± 0.10 ** & 0.10   ± 0.01 ** \\
    \text{$\text{D-TCN}_\text{Count, VADER, Tweet}$} & 0.24   ± 0.02 & 1.60   ± 0.07  & 2.59   ± 0.09 & 0.11   ± 0.01  \\
    \text{$\text{D-TCN}_\text{Count, VADER, User}$} & 0.22   ± 0.01 ** & 1.47   ± 0.07 *** & 2.39   ± 0.10 *** & 0.10   ± 0.01 *** \\
    \text{$\text{D-TCN}_\text{Count, VADER}$} & 0.29   ± 0.02 & 1.82   ± 0.08  & 2.95   ± 0.18 & 0.15   ± 0.01  \\ 
    \bottomrule
  \end{tabular}
    }
    \medskip\\
    \textbf{For D-TCNs:} One-sided Welch's t-test significances with respect to TCN. \\
\textbf{Significance codes:} 0  - ‘***’ - 0.001 - ‘**’ - 0.01 - ‘*’ - 0.05 - ‘.’ - 0.1 
\vspace{-10pt}
\end{table*}

\subsection{Experiment Setup} 

The setup is the same as in \secref{sec:exp_setup}. The model is fitted separately for each subset of the feature sets, listed in \tabref{tab:feature_sets}. Differently from \secref{sec:tcn}, we train and test each model 40 times to gather more accurate estimates of the expected performance in each case. A separate hyperparameter optimization is not conducted for the D-TCN setups to ensure comparability of the results with the baseline TCN and also with each other.

\subsection{Results} 

Here we compare the results of a single TCN (our \emph{baseline} here), analysed in \secref{sec:tcn}, to the results of the D-TCN models trained with different feature subsets from \tabref{tab:feature_sets}. The distribution of fits per each metric and model are compared with the baseline using a one-sided Welch's t-test. The results are reported in \tabref{tab:results}.

In a comparative feature analysis, it is important to consider the value of some new information in the context of a bias-variance problem. The additional information provided with each good feature set is expected to boost the performance of the model, while the variance introduced by the increased set of parameters is expected to decrease the performance. Therefore, we assess the features not only by the potential explanation they provide, but also with the noise content they possess. 

\subsubsection{Performance of individual feature sets} 
\begin{figure}[!htb]
     \includegraphics[width=0.35\textwidth]{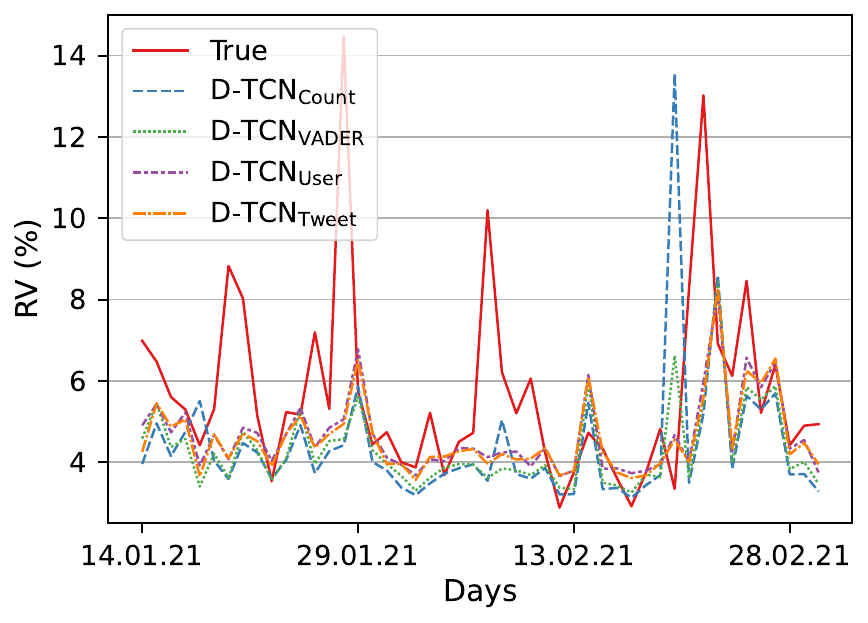}
     \caption{Mean predictions of the D-TCNs with single feature sets from \tabref{tab:feature_sets}: \emph{\textbf{Count}} in blue dashed line, \emph{\textbf{VADER}} in green dotted line, \emph{\textbf{User}} in purple dash-dot line, and \emph{\textbf{Tweet}} in orange hyphen-dot line.}
     \label{fig:single_feats}
\vspace{-10pt}
\end{figure}

We observe that only two feature sets, when used individually, perform better than the baseline TCN, namely \textit{\textbf{User}} and \textit{\textbf{Tweet}}. The former feature set both surpasses the baseline TCN performance ($p < 0.001$) and, on average of the all runs, creates the best model in our study. The \textit{\textbf{Tweet}} feature set also boosts the baseline results (with $p < 0.001$ in MAPE and MAE, $p < 0.01$ in RMSE and MSLE). In \figref{fig:single_feats}, we further observe that the predictions of \textit{\textbf{Count}} and \textit{\textbf{VADER}} feature sets have a tendency to over-value in some days, severely hurting the aggregate metrics in the displayed results. As we observe full convergence in all models during training, these artefacts can be regarded as the effects of overfitting on the training data.

\subsubsection{Performance of collections of feature sets}

The best performing models based on the significance levels and metric results were trained with the \textit{\textbf{User}} feature set. However, the feature sets \textit{\textbf{VADER}} and \textit{\textbf{Count}}, while performing poorly by themselves, experience a performance boost when combined with other feature sets. Examples of such cases are the performance obtained by the $\text{D-TCN}_\text{VADER, Tweet, User}$ and $\text{D-TCN}_\text{Count, Tweet, User}$. Although individually the \textit{\textbf{Tweet}} and \textit{\textbf{User}} sets are superior to others, when combined they benefit from the addition of \textit{\textbf{VADER}} in MAE metric, and \textit{\textbf{Count}} in RMSE and MSLE metrics. These performance changes, albeit not very significant, might be indications that with proper architectures, the interaction effects between the metadata information and other data sources such as semantic content and communication volume can be utilized to improve the results in volatility prediction.

\begin{figure}[!bth]

\begin{minipage}[b]{0.45\linewidth}
  \centering
  \centerline{\includegraphics[width=\linewidth]{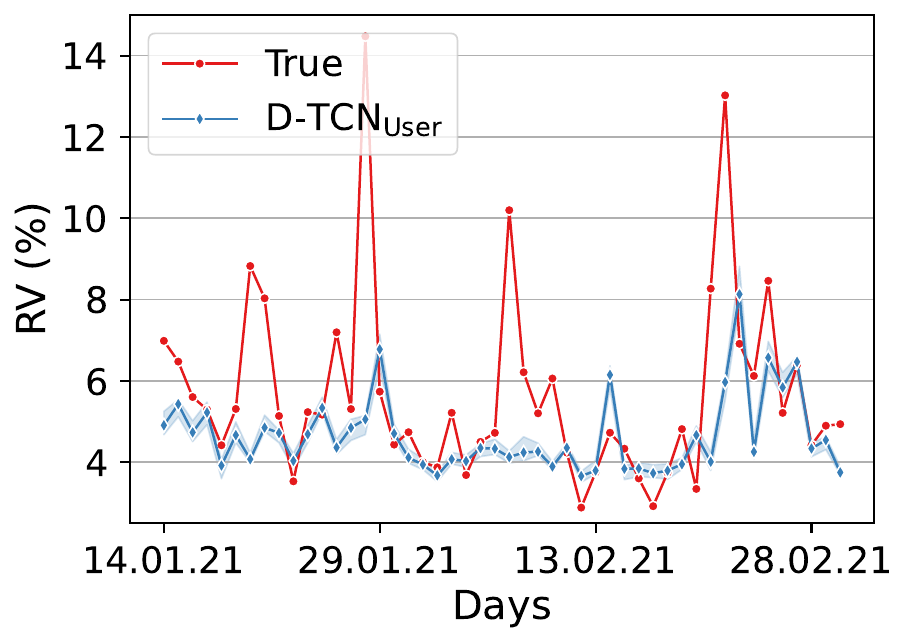}}
  \centerline{(a) $\text{D-TCN}_\text{User}$}
\end{minipage}
\hfill
\begin{minipage}[b]{0.45\linewidth}
  \centering
  \centerline{\includegraphics[width=\linewidth]{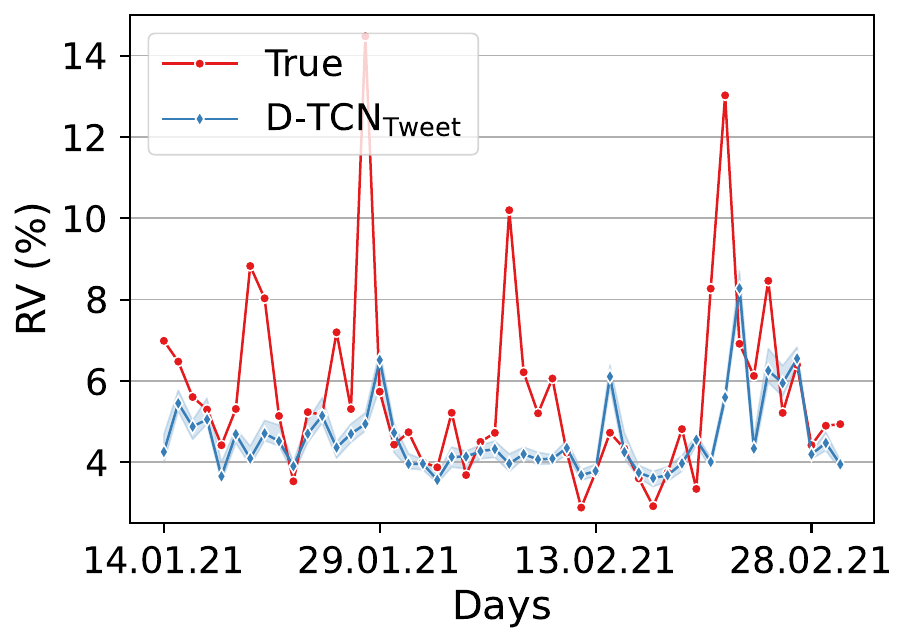}}
  \centerline{(b) $\text{D-TCN}_\text{Tweet}$}
\end{minipage}
\bigskip
\\ 
\begin{minipage}[b]{0.45\linewidth}
  \centering
  \centerline{\includegraphics[width=\linewidth]{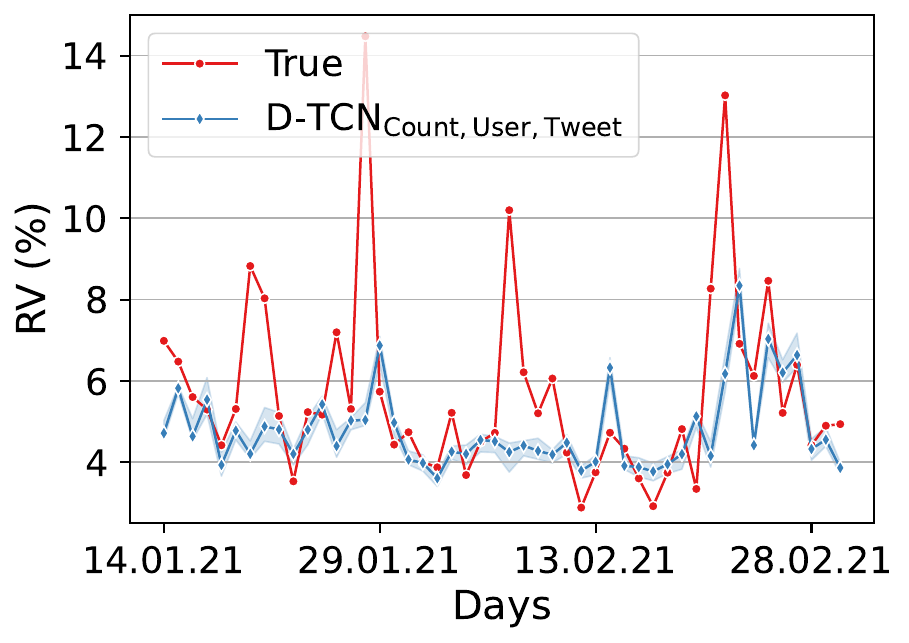}}
  \centerline{(c) $\text{D-TCN}_\text{Count, User, Tweet}$}
\end{minipage}
\hfill
\begin{minipage}[b]{0.45\linewidth}
  \centering
  \centerline{\includegraphics[width=\linewidth]{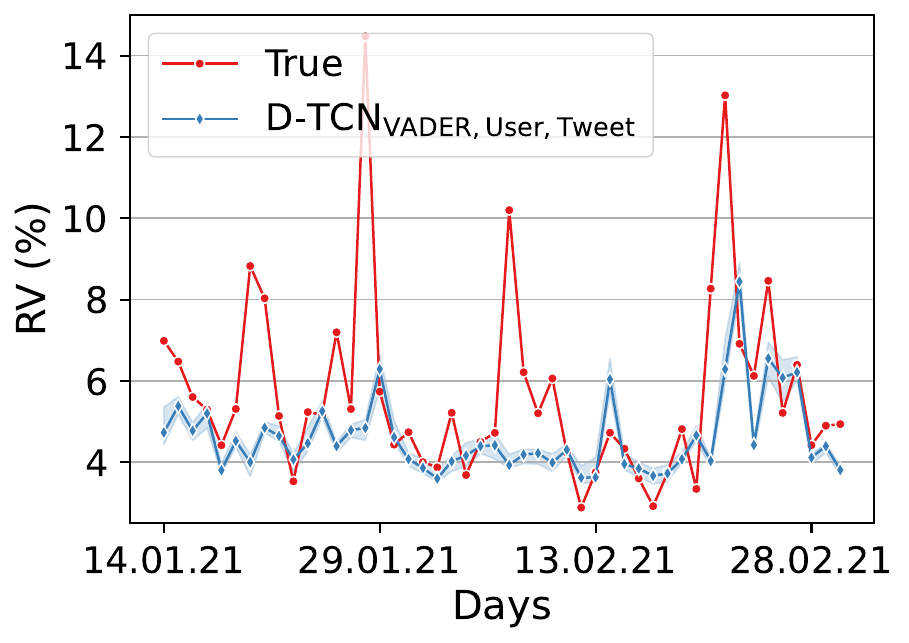}}
  \centerline{(d) $\text{D-TCN}_\text{VADER, User, Tweet}$}
\end{minipage}

\caption{Predicted vs. true realized volatilities on test sets. Shaded regions stand for the 95\% confidence intervals, estimated via bootstrapping (using 1000 bootstrap samples).}
\label{fig:final_plots}
\vspace{-10pt}
\end{figure}

\subsubsection{Performances of the best combinations at different label quartiles} 
\begin{figure}[!thb]

\begin{minipage}[b]{0.45\linewidth}
  \centering
  \centerline{\includegraphics[width=\linewidth]{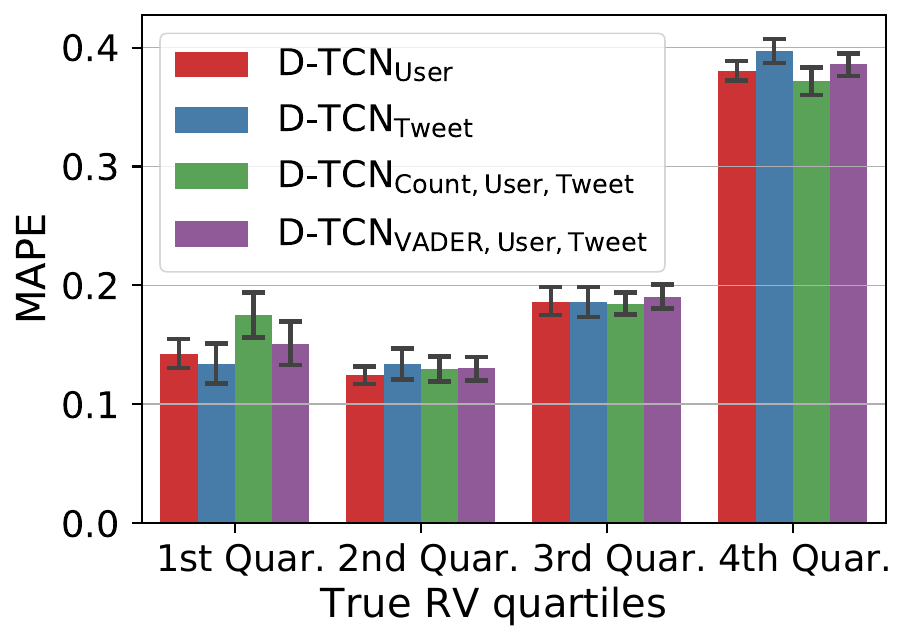}}
  \centerline{(a) Best $\text{D-TCN}$ models}
\end{minipage}
\hfill
\begin{minipage}[b]{0.45\linewidth}
  \centering
  \centerline{\includegraphics[width=\linewidth]{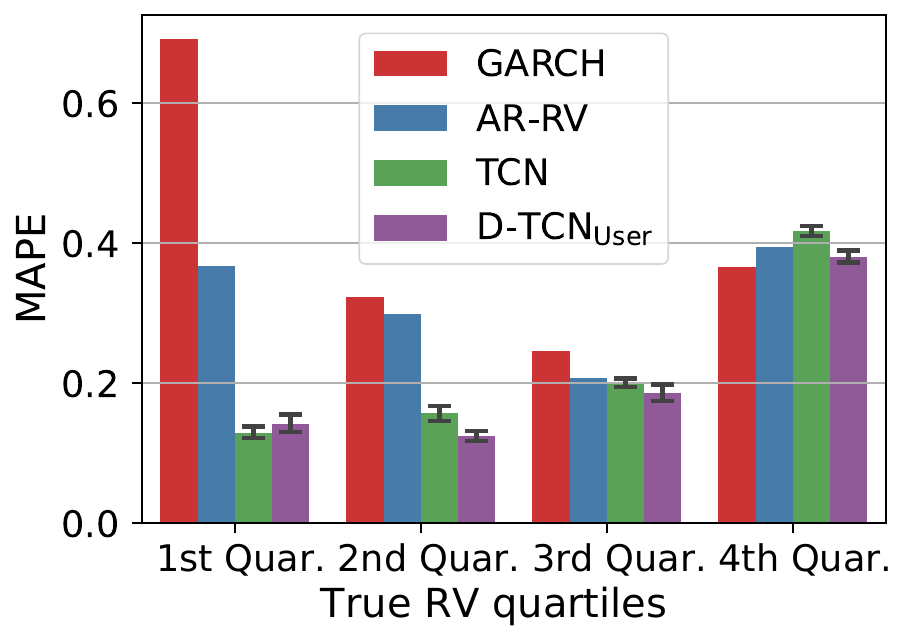}}
  \centerline{(b) Classical vs. deep models}
\end{minipage}

\ifarxiv
\vspace{-4pt}
\fi
 \caption{Mean MAPE values of each model for the given quartile of the true labels. Bands stand for 95\% confidence intervals.}
 \label{fig:quar_errors}
\ifarxiv
\vspace{-18pt}
\else
\vspace{-10pt}
\fi
\end{figure}

Since performance metrics relate to an average achieved performance across the full dataset, we further look into behaviours of D-TCN models for different target volatilities. In \figref{fig:quar_errors}, we observe that despite the good performance of the \textit{\textbf{User}} feature set in the aggregate metrics, it does not provide absolute performance boost in all target quartiles: \textit{\textbf{Tweet}} feature set is more useful in the lowest quartiles, while the addition of \textit{\textbf{Count}} appears to improve the MAPE metric in higher RV periods. Tweet volume has been previously connected with high volatility in Bitcoin prices~\cite{count_and_volatility}, and our results corroborate with this insight. Furthermore, when compared to TCN, $\text{D-TCN}_\text{User}$ has better performance in the second, third, and fourth quartiles of the true RV, conveying again that peaks are more precisely forecasted with the introduction of user information.

\section{Conclusion}\label{sec:discuss}

In this study, our aim was threefold: (i) to analyse the deep learning-based  alternatives to standard econometric models used for realized volatility forecasting; (ii) to design and develop a deep learning architecture for realized volatility forecasting that can utilize social media data; and (iii) to understand the social media factors that help in predicting realized volatility. 

To achieve our goals, we employed Twitter API in a \emph{stream-watcher} fashion, so that the tweets which contained Bitcoin-related strings were saved instantly. We then utilized the obtained textual and other accompanying information of Bitcoin-related tweets and used it together with intraday Bitcoin price data in the prediction task. The gathered data was first preprocessed by pruning and refactoring, so that all social media data entries had the same format, and contained only the information that could potentially aid the realized volatility forecast.

The double-TCN architectural design was developed via a thorough three-stage analysis. First, by comparing several deep learning architectures (namely LSTM, GRU, and TCN) to econometric baseline models, we observed that TCN yields the most promising results. Following this step, we built a two-level TCN model that allows for concurrent and modular use of log-returns and information obtained from social media. Finally, to assess the influence of the information added, we trained the D-TCN with combinations of different feature sets and compared the results obtained from each combination with the TCN which was trained using only the log-returns.

Overall, our experiments produced an unexpected outcome: somewhat counter-intuitively, semantic content of the tweet's text is \emph{not} the most informative for the predictions of realized volatility. On the contrary, information about authors of the tweets (e.g., follower counts and friend counts) convey much more predictive power. Feature sets that combined \textit{\textbf{User}} information and \textit{\textbf{Tweet}} information together with information about the total tweet volume (\textit{\textbf{Count}}) were more predictive than \textit{\textbf{Count}} would be on its own, particularly in high realized volatility regimes.

In the future, we believe that a higher prediction accuracy can be achieved by forming a dataset that additionally contains interaction statistics for individual tweets  such as retweet count, favourite count and reply count. Incorporating a complex embedding structure like BERT \cite{devlin2019bert} instead of VADER, and making use of the network-based structure of Twitter via Graph Neural Networks may also improve the precision of realized volatility forecasting.

\ifarxiv
\vspace{-13pt}
\fi
\begin{acks}
M.E.A., M.E. and K.K. thank Prof.\ Dr.\ Ce Zhang for their help during this research, Benjamin Suter for his help on the collected tweet dataset, and the the ETH Z\"urich DS3Lab for giving us access to their computer infrastructure. N.A.F. and V.V. are supported by the European Union - Horizon 2020 Program under the scheme ‘INFRAIA-01-2018-2019 - Integrating Activities for Advanced Communities’, Grant Agreement no. 871042, ‘SoBigData++: European Integrated Infrastructure for Social Mining and Big Data Analytics’ (http://www.sobigdata.eu).
\end{acks}

\ifarxiv
\bibliographystyle{unsrt}
\bibliography{bibliography}
\else
\bibliographystyle{unsrt}
\balance
\bibliography{bibliography}
\fi

\ifarxiv
\appendix

\section{Metrics}
\label{app_metrics}
The following metrics are defined in terms of i) $y_i$, the actual realized volatility on the $i$\textsuperscript{th} day, and ii) $\hat{y}_i$, the predicted realized volatility on the $i$\textsuperscript{th} day. $N$ denotes the total number of predicted datapoints in a given horizon.

\begin{align}
\textrm{MAPE} = \frac{1}{N} \sum_{i=1}^{N} \left|{\frac{y_i - \Hat{y}_i} {y_i}}\right| \end{align}

\begin{align}
\textrm{MAE} = \frac{1}{N} \sum_{i=1}^{N} |{y_i - \Hat{y}_i}| \
\end{align}

\begin{align}
\textrm{RMSE} = \sqrt{\frac{1}{N} \sum_{i=1}^{N} (y_i - \Hat{y}_i)^2} 
\end{align}

\begin{align}
\textrm{MSLE} = \frac{1}{N} \sum_{i=1}^{N} (\log(y_i + 1) - \log(\Hat{y}_i + 1))^2
\end{align}

\section{Additional Results}
In \secref{sec:twitter_tcn}, the individual feature sets \textit{\textbf{VADER}}, \textit{\textbf{Count}}, \textit{\textbf{User}} and \textit{\textbf{Tweet}} were presented at \figref{fig:single_feats}, and best performing four combinations were presented at \figref{fig:final_plots}, leaving the results for 9 feature set combinations unseen. For completeness, results for the remaining 9 feature set combination plots can be observed in Figures \ref{fig:add_res_plot} and \ref{fig:add_res_plot2}.

\begin{figure}[H]
\begin{minipage}[b]{.48\linewidth}
  \centering
  \centerline{\includegraphics[width=\linewidth]{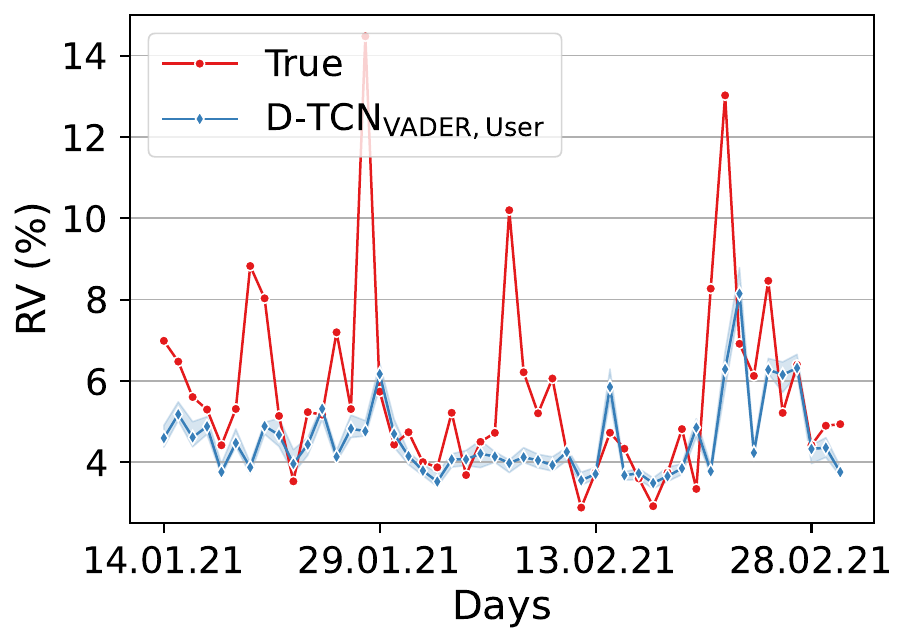}}
  \centerline{(a) $\text{D-TCN}_\text{VADER, User}$}
\end{minipage}
\hfill
\begin{minipage}[b]{0.48\linewidth}
  \centering
  \centerline{\includegraphics[width=\linewidth]{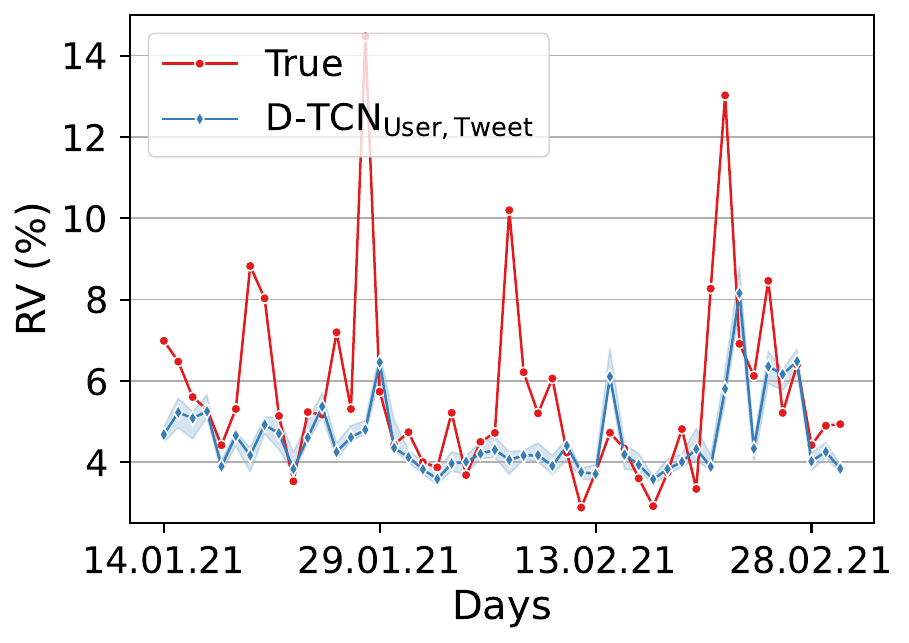}}
  \centerline{(b) $\text{D-TCN}_\text{User, Tweet}$}
\end{minipage}
\bigskip
\\ 
\begin{minipage}[b]{.48\linewidth}
  \centering
  \centerline{\includegraphics[width=\linewidth]{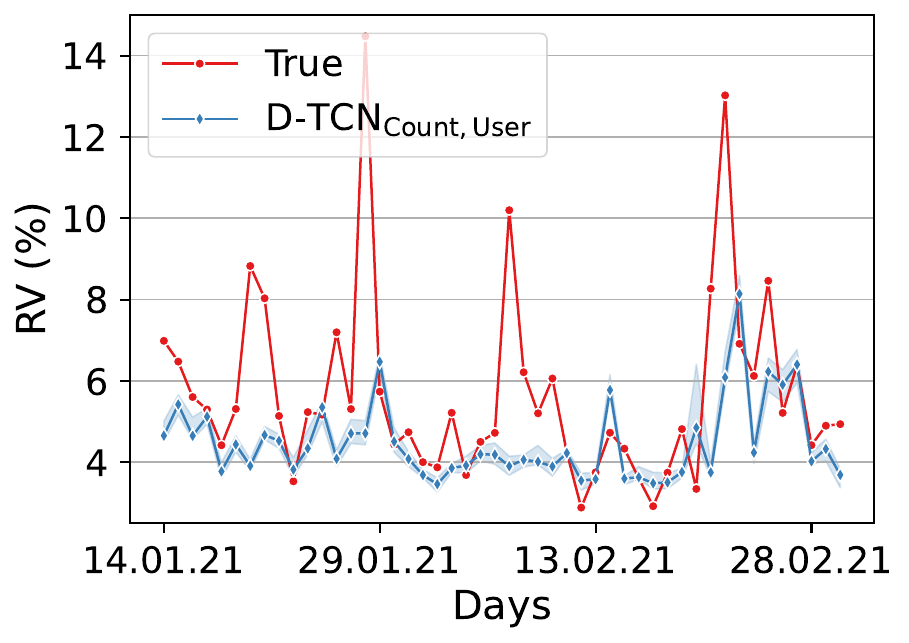}}
  \centerline{(c) $\text{D-TCN}_\text{Count, User}$}
\end{minipage}
\hfill
\begin{minipage}[b]{0.48\linewidth}
  \centering
  \centerline{\includegraphics[width=\linewidth]{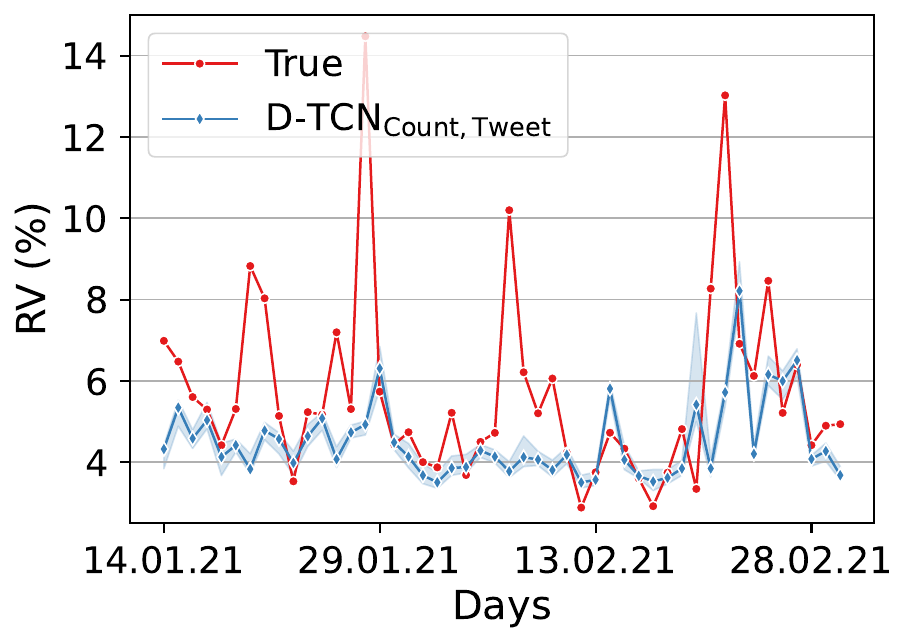}}
  \centerline{(d) $\text{D-TCN}_\text{Count, Tweet}$}
\end{minipage}

\caption{Results of feature set combinations not presented in \secref{sec:twitter_tcn}.}
\label{fig:add_res_plot}
\vspace{-10pt}
\end{figure}

\begin{figure}[H]
\begin{minipage}[b]{0.48\linewidth}
  \centering
  \centerline{\includegraphics[width=\linewidth]{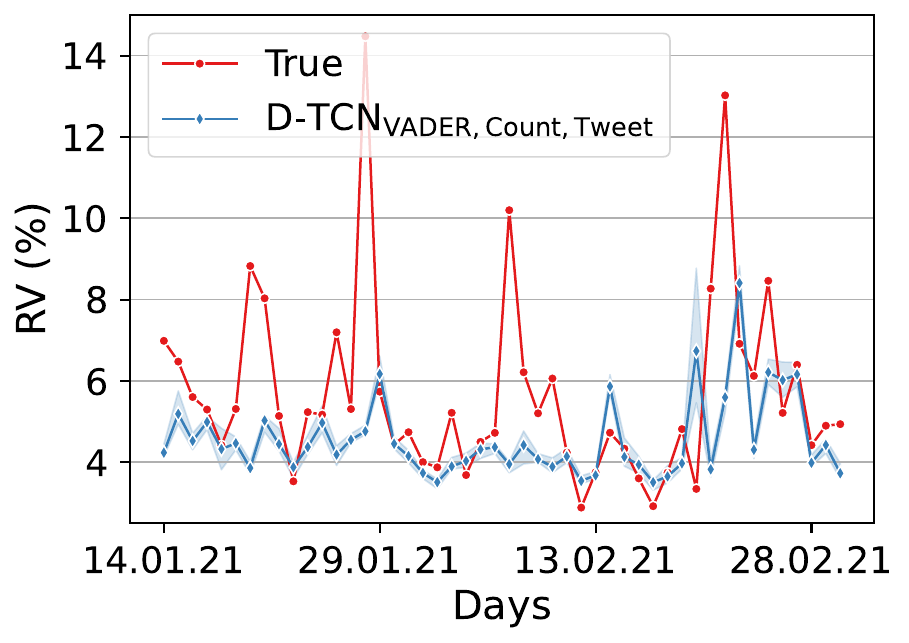}}
 \centerline{(e) $\text{D-TCN}_\text{VADER,Count, Tweet}$}
\end{minipage}
\begin{minipage}[b]{0.48\linewidth}
  \centering
  \centerline{\includegraphics[width=\linewidth]{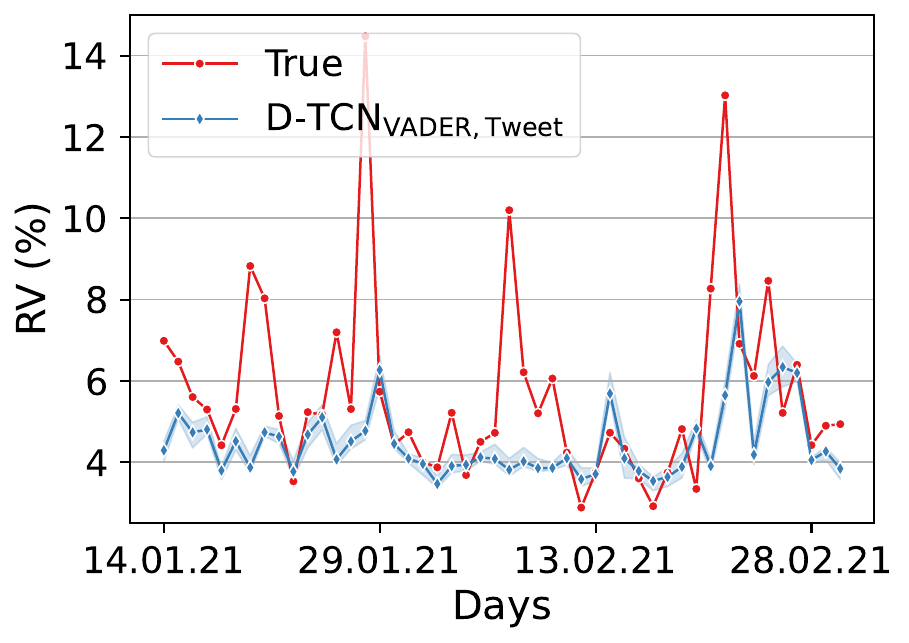}}
  \centerline{(f) $\text{D-TCN}_\text{VADER, Tweet}$}
\end{minipage}
\bigskip\\
\begin{minipage}[b]{0.48\linewidth}
  \centering
  \centerline{\includegraphics[width=\linewidth]{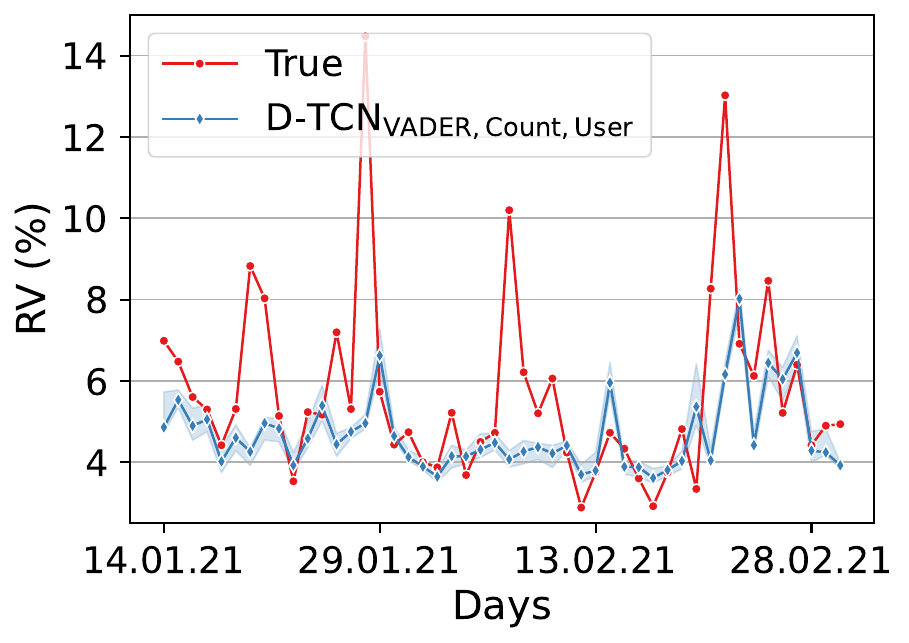}}
 \centerline{(g) $\text{D-TCN}_\text{VADER, Count, User}$}
\end{minipage}
\begin{minipage}[b]{0.48\linewidth}
  \centering
  \centerline{\includegraphics[width=\linewidth]{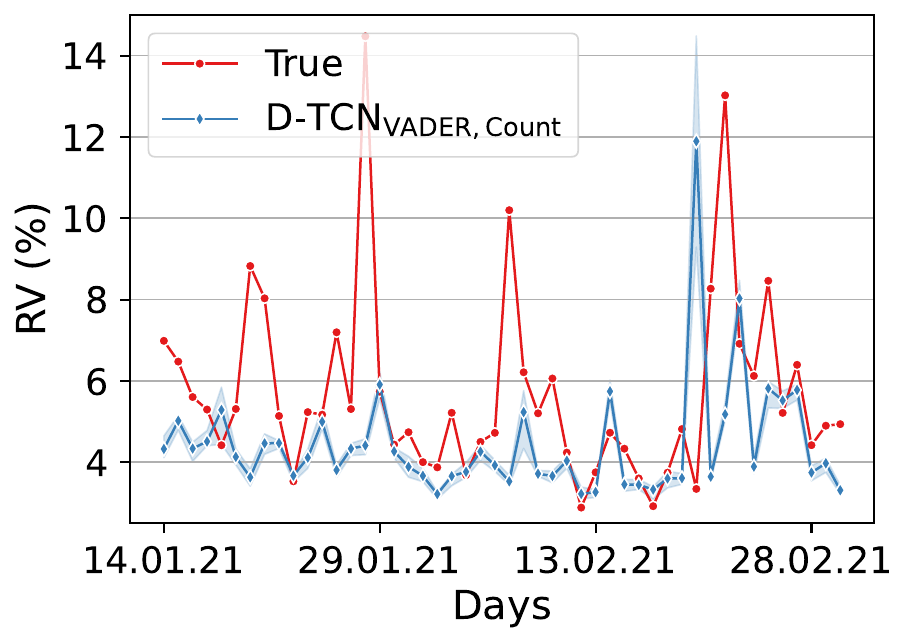}}
  \centerline{(h) $\text{D-TCN}_\text{VADER, Count}$}
\end{minipage}
\bigskip\\
\begin{minipage}[b]{0.48\linewidth}
  \centering
  \centerline{\includegraphics[width=\linewidth]{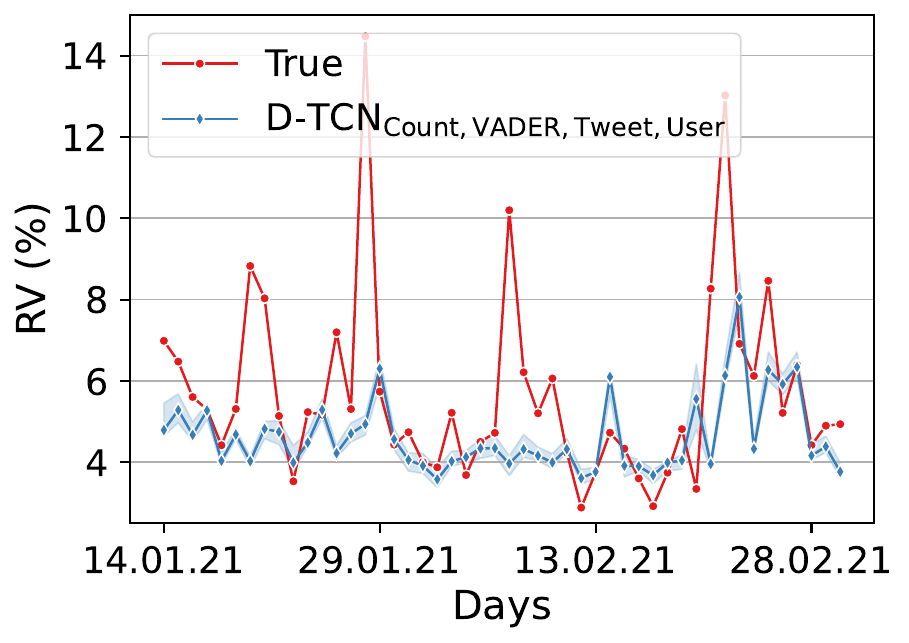}}
  \centerline{(i) $\text{D-TCN}_\text{Count, VADER, Tweet, User}$}
  \end{minipage}
  
\caption{Continuation of results of feature set combinations not presented in \secref{sec:twitter_tcn}.}
\label{fig:add_res_plot2}
\end{figure}

\begin{table}[!bth]
  \caption{Hyperparameter search space for LSTM and GRU}
  \label{tab:hpo_lstm}
  \begin{tabular}{lllll}
    \toprule
     \text{Parameter} &\text{Low} &\text{High} &\text{Best LSTM} & \text{Best GRU} \\
    \midrule
    \text{Recurrent Dimension} & 32 & 512 & 261 & 43  \\
    \text{Dropout} & 0 & 0.5 & 0.0237 & 0.0544 \\
    \text{$\epsilon$ (Loss epsilon)} & 0.01  & 0.1 & 0.0364 & 0.0431  \\
    \text{Learning Rate} & $10^{-7}$ & $10^{-2}$ & 0.00521 & 0.00973 \\
    \text{Weight Decay} & $10^{-9}$ & $10^{-2}$ & $9.25\times10^{-7}$ & $1.59\times10^{-8}$\\
    \bottomrule
  \end{tabular}
\vspace{-10pt}
\end{table}

\begin{table}[!hbt]
  \caption{Hyperparameter search space for TCN}
  \label{tab:hpo_tcn}
  \begin{tabular}{ll p{2cm} l}
    \toprule
     \text{Parameter} &\text{Low} &\text{High} &\text{Best} \\
    \midrule
    \text{Filter count} & 32  & 512 & 287\\
    \text{Dropout} & 0 & 0.5 & 0.217\\
    \text{$\epsilon$ (Loss epsilon)} & 0.01  & 0.1 & 0.0913 \\
    \text{Learning Rate} & $10^{-7}$ & $10^{-2}$ & $6.49\times10^{-5}$ \\
    \text{Weight Decay} & $10^{-9}$ & $10^{-2}$ & $5.93\times10^{-6}$ \\
    \text{Kernel Size} & 2 & 6 & 5 \\
    \text{Dilation rate} & 2 & 4 & 4 \\
    \text{Skip Connections} & False & True & True \\
    \text{Normalization} & None & One of \textit{Batch}, \textit{Layer} or \textit{Weight} & None \\
    \bottomrule
    \\
  \end{tabular}
\vspace{-10pt}
\end{table}

\begin{figure}[!bt]
\begin{minipage}[b]{0.48\linewidth}
  \centering
  \centerline{\includegraphics[width=\linewidth]{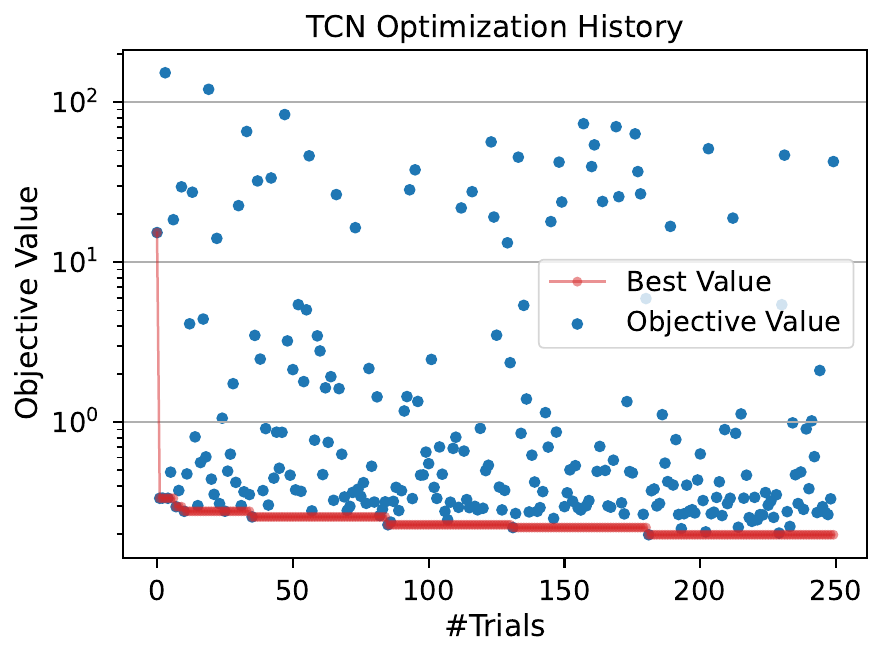}}
  \centerline{(a) TCN}
\end{minipage}
\begin{minipage}[b]{0.5\linewidth}
  \centering
  \centerline{\includegraphics[width=\linewidth]{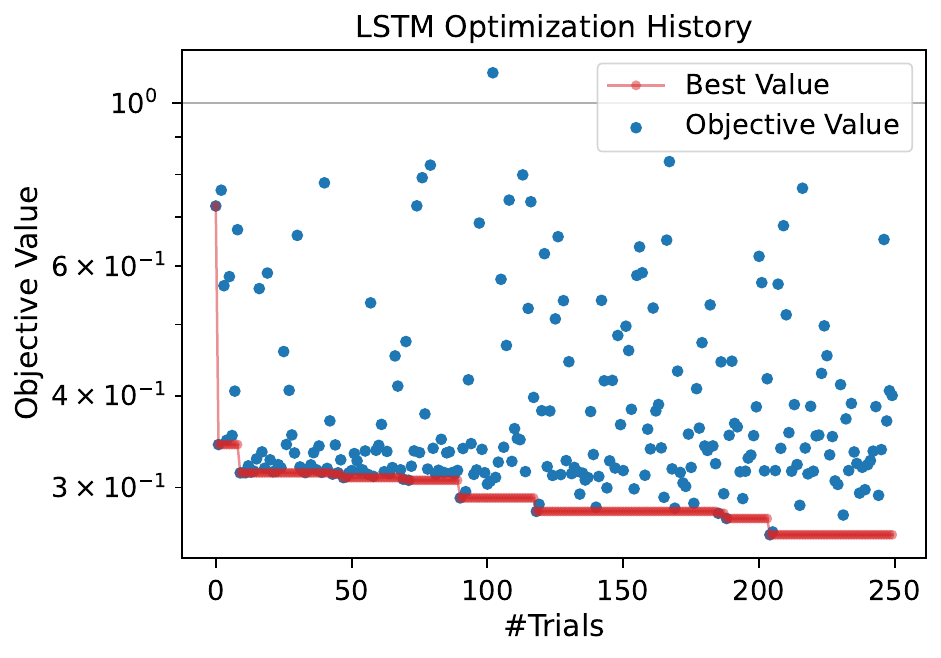}}
  \centerline{(b) LSTM}
\end{minipage}
\bigskip
\\ 
\begin{minipage}[b]{0.48\linewidth}
  \centering
  \centerline{\includegraphics[width=\linewidth]{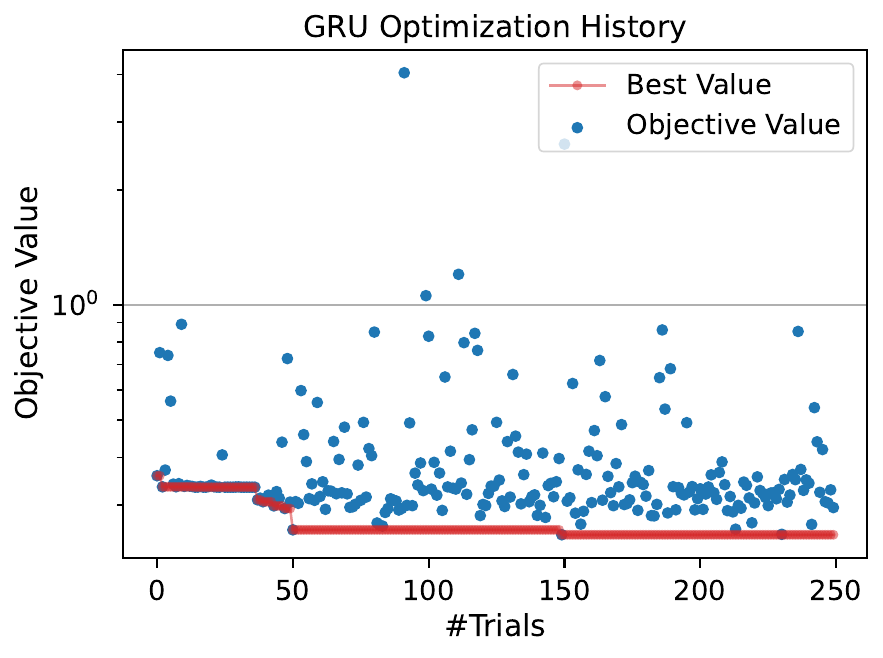}}
  \centerline{(c) GRU}
\end{minipage}
\caption{Trial-error plots for TCN, LSTM and GRU}
\label{fig:hpo_hist}
\vspace{-10pt}
\end{figure}

\begin{figure}[!bt]
\begin{minipage}[b]{0.48\linewidth}
  \centering
  \centerline{\includegraphics[width=\linewidth]{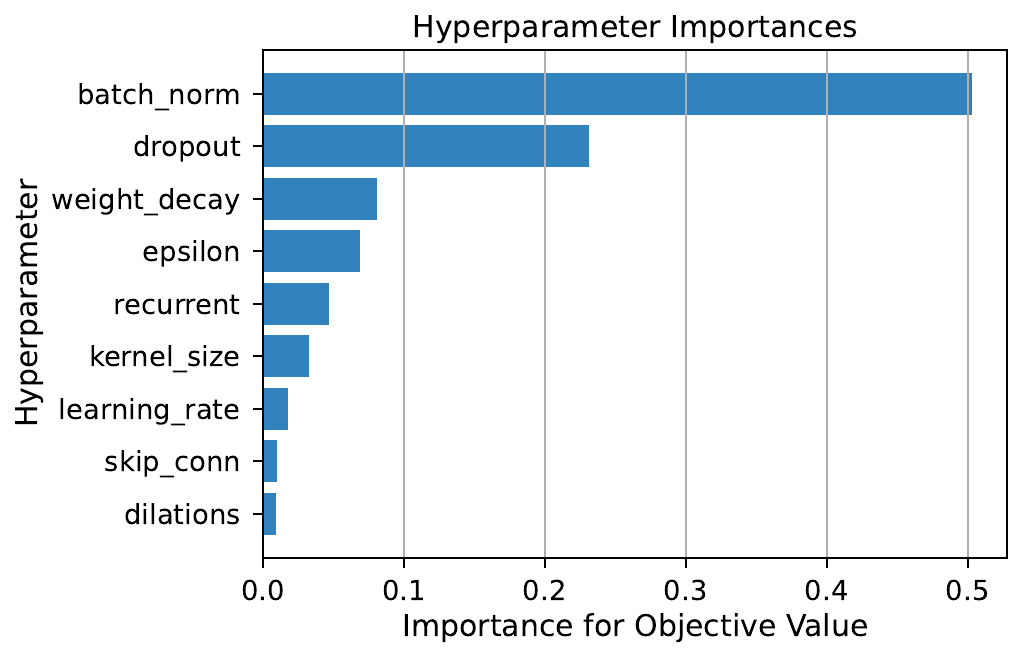}}
  \centerline{(a) TCN}
\end{minipage}
\begin{minipage}[b]{0.48\linewidth}
  \centering
  \centerline{\includegraphics[width=\linewidth]{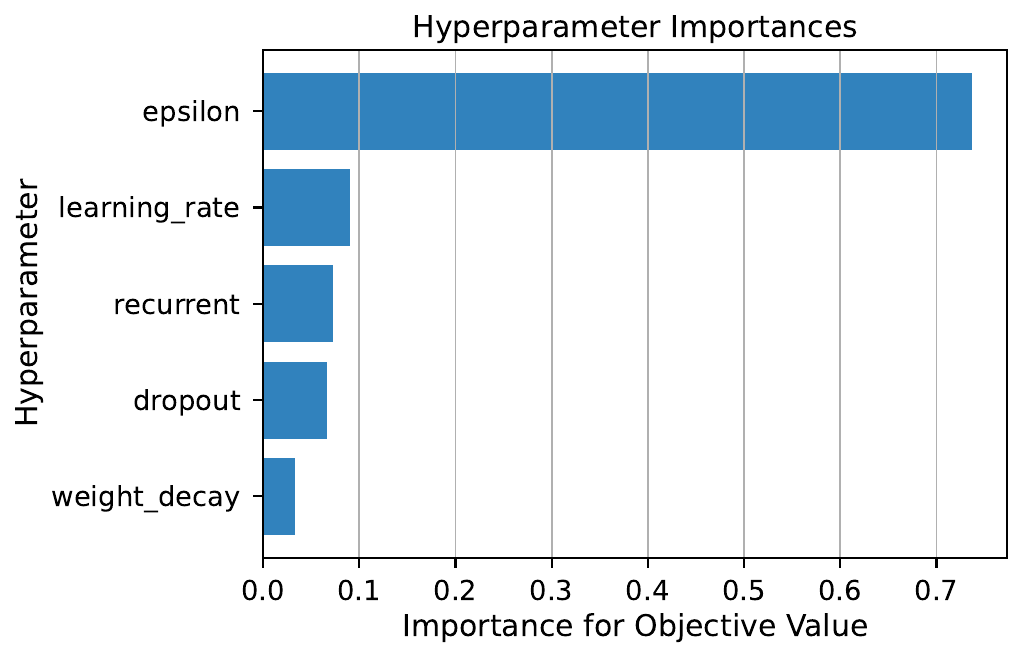}}
  \centerline{(b) LSTM}
\end{minipage}
\bigskip
\\ 
\begin{minipage}[b]{0.48\linewidth}
  \centering
  \centerline{\includegraphics[width=\linewidth]{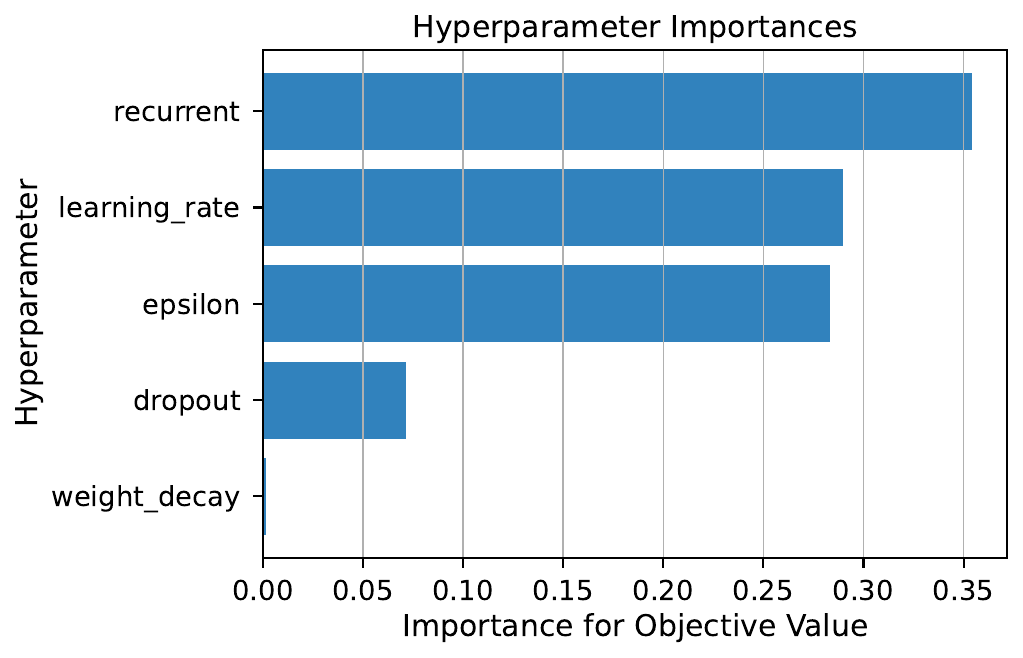}}
  \centerline{(c) GRU}
\end{minipage}

\caption{Importance charts of hyperparameters for TCN, LSTM and GRU}
\label{fig:hpo_importance}

\vspace{-10pt}
\end{figure}

\newpage

\section{Hyperparameter optimization}
\label{app_hpo}

Python library \textit{optuna} \cite{optuna} is used to conduct hyperparameter optimization for all three baseline deep learning models, namely LSTM, GRU, and TCN, as well as D-TCN. A multivariate Tree-structured Parzen Estimator sampler (TPE)~\cite{tpe} is used in a Bayesian setting to minimize the MAPE error on a training set of 72 days and a validation set of 24 days (from the 96 days that is initially determined as the full training set of the experiments). The epoch count is kept constant at 30. All models are optimized for 250 iterations. 

The parameter search spaces and the best parameter values for LSTM/GRU and TCN are shown in \tabref{tab:hpo_lstm} and \tabref{tab:hpo_tcn} respectively. \figref{fig:hpo_hist} demonstrate the error versus trial count plot for each model. \figref{fig:hpo_importance} show the importance scores with respect to the objective value (MAPE). As depicted in \figref{app_hpo}, LSTM achieves 0.2588 MAPE on the validation set at the best trial, GRU scores 0.2513, and TCN 0.1977.

\fi

\end{document}